\documentclass[twocolumn]{aastex62}

\usepackage{epsfig}
\usepackage{epstopdf}
\usepackage{graphics}
\usepackage{amsmath}
\usepackage{xcolor}
\usepackage{aas_macros}

\usepackage{array}
\usepackage{booktabs}
\usepackage{amsmath,amssymb,amsfonts,amsbsy}
\usepackage{mathrsfs}
\usepackage{url}
\usepackage{multirow}
\usepackage{xcolor}
\usepackage{color}
\usepackage{ulem}
\usepackage{enumerate}
\usepackage{lineno}

\renewcommand{\vec}[1]{\boldsymbol{#1}}

\renewcommand{\figurename}{Fig.}
\renewcommand{\tablename}{Tab.}
\begin{document}
%\linenumbers

\title{Statistical study of the optimal local sources for cosmic ray nuclei and electrons}

\correspondingauthor{ Bing-qiang Qiao, Wei Liu, Shu-wang Cui,\\
	Yi-qing Guo}
\email{ qiaobq@ihep.ac.cn, liuwei@ihep.ac.cn, cuisw@hebtu.edu.cn,\\
	guoyq@ihep.ac.cn}

\author{Qing Luo}
\affiliation{
Hebei Normal University, Shijiazhuang, Hebei 050024, China
}
\affiliation{Key Laboratory of Particle Astrophysics,
	Institute of High Energy Physics, Chinese Academy of Sciences, Beijing 100049, China
}

\author{Bing-qiang Qiao}
\affiliation{Key Laboratory of Particle Astrophysics,
	Institute of High Energy Physics, Chinese Academy of Sciences, Beijing 100049, China
}

\author{Wei Liu}
\affiliation{Key Laboratory of Particle Astrophysics,
Institute of High Energy Physics, Chinese Academy of Sciences, Beijing 100049, China
}

%\author{Qiang Yuan}
%\affiliation{Key Laboratory of Dark Matter and Space Astronomy,
%	Purple Mountain Observatory, Chinese Academy of Sciences,
%	Nanjing 210008, China
%}
%\affiliation{School of Astronomy and Space Science, University of Science and Technology of China,
%	Hefei 230026, China
%}
%%\affiliation{Center for High Energy Physics, Peking University,
%%	Beijing 100871, China}
%

\author{Shu-wang Cui}
\affiliation{
Hebei Normal University, Shijiazhuang, Hebei 050024, China
}
\author{Yi-qing Guo}
\affiliation{Key Laboratory of Particle Astrophysics,
Institute of High Energy Physics, Chinese Academy of Sciences, Beijing 100049, China
}
\affiliation{University of Chinese Academy of Sciences, Beijing 100049, China
}

\begin{abstract}
{%It is generally believed that the supernova remnant (SNR) can simultaneously accelerated the nuclei and electron to very high energy.
The local sources, such as the Geminga supernova remnant (SNR), may play an important role in the anomaly of protons, electrons and anisotropy in the past works. In fact, there exist twelve SNRs around the solar system within $1$ kiloparsec (kpc). One question is that can other SNRs also possibly contribute to the spectra of nuclei and electrons and explain the special structure of the anisotropy? In this work, under the spatial-dependent propagation, we systematically study the contribution of all local SNRs within 1 kpc around the solar system to the spectra of
nuclei and electrons, as well as the energy dependence of the anisotropy. As a result, only the Geminga, the Monogem, and the Vela SNRs have quantitative contribution to the
nuclei and electron spectra and the anisotropy. Here, the Geminga SNR is the sole optimal candidate and the
Monogem SNR is controversial due to the tension of the anisotropy between the model calculation and the observations. The Vela SNR contributes to a new spectral structure beyond TeV energy, hinted by the HESS, the VERITAS, the DAMPE, and the CALET measurements. More interestingly, the electron anisotropy
satisfies the Fermi-LAT limit below TeV energy, but rises greatly and reaches $10\%$ at several
TeV. This novel structure will shed new light on verifying our model. We hope that the new structure of the electron spectrum and anisotropy can be observed by the space-borne DAMPE and HERD, and the ground-based HAWC and LHAASO experiments in the near future.
}
\end{abstract}

\section{Introduction}

  It is well known that supernova remnants (SNRs) are the dominant sources of
  galactic cosmic rays (GCRs) \citep{1989ApJ...342..807B,2013A&ARv..21...70B,2013Sci...339..807A}. In this scenario, the expanding diffusive shocks accelerate cosmic rays (CRs) to very high energy (VHE). The electrons and nuclei are concomitant and can both be accelerated to VHE simultaneously. This means that the nuclei and electrons should have some common origin \citep{2013PhLB..727....1Y}. The combined study for such multi-messenger topic is important to unveil the enigma of the origins of the CRs. And the typical properties are required to be observed to support this view of point.

  The measurements of CRs are stepping into a high precision era with the new generation of
  space-borne and ground-based experiments. A series of new phenomena are revealed by those
  precise measurements. Firstly, a fine structure of spectral hardening at 200 GeV for nuclei
  was discovered by the ATIC-2, the CREAM and the PAMELA experiments \citep{2007BRASP..71..494P,2009BRASP..73..564P,2010ApJ...714L..89A,2017ApJ...839....5Y,2011Sci...332...69A}. Lately, the AMS-02 experiment confirmed it with unprecedented precision \citep{2015PhRvL.114q1103A,2015PhRvL.115u1101A}. More interestingly, the spectral break-off around $\sim$14 TeV was observed by the CREAM, the NUCLEON and the DAMPE experiments \citep{2017ApJ...839....5Y,2017JCAP...07..020A,2018JETPL.108....5A,2019SciA....5.3793A,2021PhRvL.126t1102A}.
  Three categories of models such as the local sources model \citep{2013APh....50...33S,2017PhRvD..96b3006L,2019JCAP...10..010L,2012ApJ...752...68V}, the combined effects from
  different group sources model \citep{2020arXiv201002826M} and the spatial-dependent propagation (SDP) model \citep{2016ApJ...819...54G,2016ChPhC..40a5101J,2018PhRvD..97f3008G,2018ApJ...869..176L} are proposed to explain the these new structures. Similar to that of the nuclei, the electrons should also exist such a component of spectral break-off above $\sim$100 GeV. The recent precise spectrum measurement shows sharp drop-off at 284 GeV for positrons, but
  no obvious change until TeV for the total spectrum of electrons and positrons \citep{2019PhRvL.122d1102A}.
  The deficit of positrons above 284 GeV requires the compensation of primary electrons.
  This means that the excess of electrons exists in a way similar to the fine structure of nuclei, which is
  possibly accelerated by the SNRs scenario \citep{2011JCAP...02..031M, 2012A&A...544A..92B,2012APh....39....2S, 2014JCAP...04..006D,2018ApJ...863...30F, 2019MNRAS.484.3491T}.

  What more important is the anisotropy evolution with energy. Though the arrival directions of GCRs are highly isotropic due to their diffusive propagation in the galactic magnetic field, a weak dipole-like anisotropy is consistently observed, with difference in intensity of up to $\sim 10^{-4}-10^{-3}$.
  So far, a large amount of observations with anisotropies ranging from TeV to PeV have been carried out by the ground-based experiments, for example, the Super-Kamiokande \citep{2007PhRvD..75f2003G}, the Tibet \citep{2005ApJ...626L..29A, 2006Sci...314..439A, 2010ApJ...711..119A, 2017ApJ...836..153A}, the Milagro \citep{2008PhRvL.101v1101A, 2009ApJ...698.2121A}, the IceCube/Ice-Top \citep{2010ApJ...718L.194A, 2011ApJ...740...16A, 2012ApJ...746...33A, 2013ApJ...765...55A, 2016ApJ...826..220A}, the ARGO-YBJ \citep{2013PhRvD..88h2001B, 2015ApJ...809...90B} and the HAWC \citep{2014ApJ...796..108A}. It is clear that the phase of the anisotropy below $100$ TeV energy roughly directs to the galactic anticenter, which is totally paradoxical with the conventional propagation model (CPM). However, above $100$ TeV, the phase gradually turns to the direction of the galactic center, until about 100 PeV energies. This is consistent with the
  expectation of the CPM. Correspondingly, the amplitude has experienced similar transition at the critical energy of $100$ TeV. In addition, what most important is that there exists a common transition energy scale between the structures of the energy spectra and the anisotropy. The local sources possibly play a very important role in
  resolving the conjunct problems of the spectra and the anisotropy \citep{2019JCAP...10..010L,2019JCAP...12..007Q}.

\begin{figure}[!htb]
	\centering
	\includegraphics[height=7.3cm, angle=0]{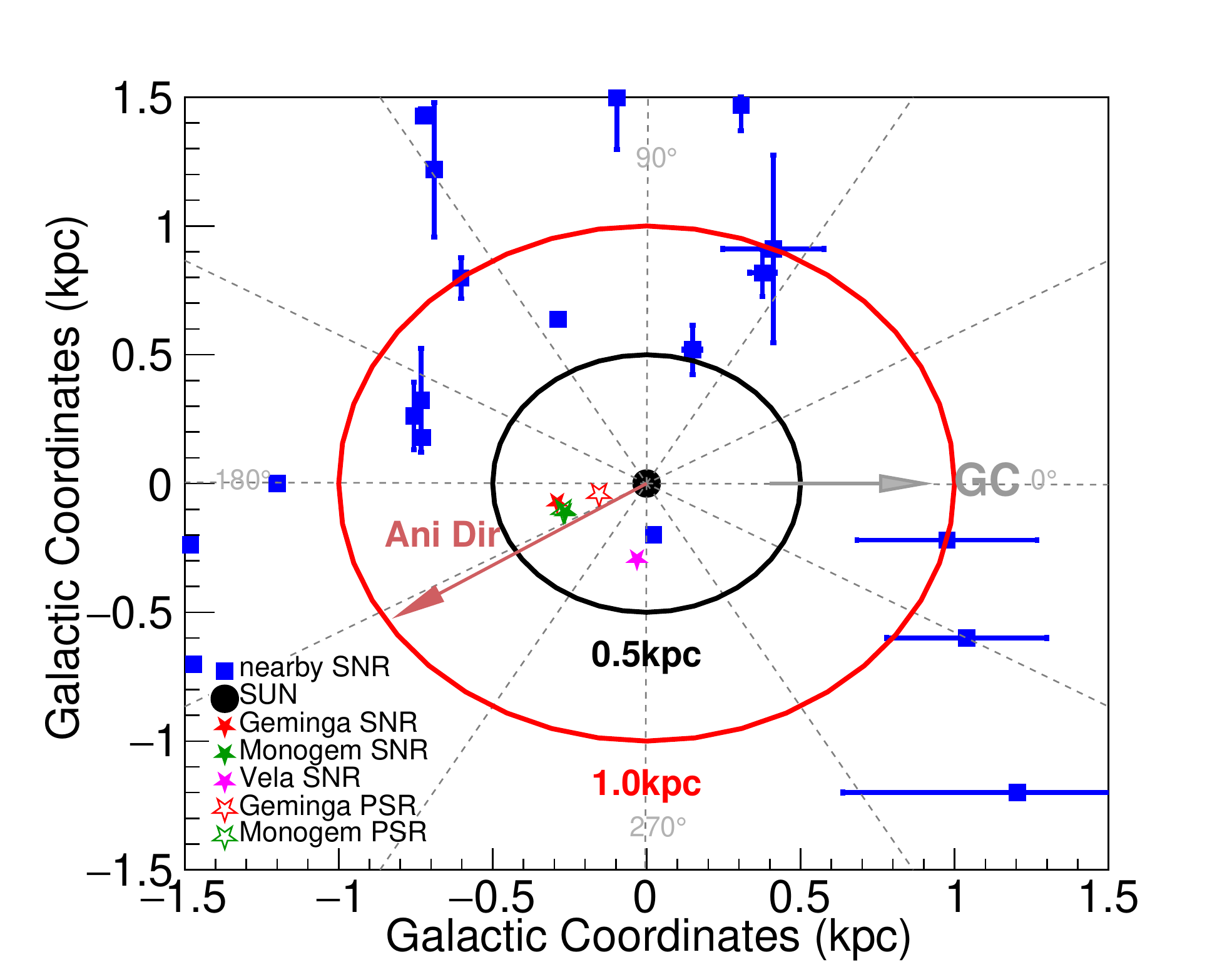}
	\caption{Face-on view of the galaxy showing positions of the local SNRs within $1$ kpc from the solar system. The blue solid squares mark the SNRs. The solide circle at the center represents the location of the solar system. The solid red, green and pink pentagrams represent the Geminga, the Monogem, and the Vela SNRs, respectively, and the hollow ones represent the homologous pulsars. The gray arrow points to the galactic center and the brown arrow indicates the direction of CRs anisotropy below $\sim 100$ TeV.
	}
	\label{fig:local_sources}
\end{figure}

  Based on the above discussion, the local sources are required to reproduce the multi-messenger anomaly for the spectra of protons and electrons and the nuclear anisotropy.
  In our recent work, we propose a local source under the SDP model to reproduce the co-evolution of the spectra and the anisotropy. We find that the Geminga SNR at its birth place could be a preferred candidate \citep{2019JCAP...10..010L,2019JCAP...12..007Q,2021JCAP...05..012Z}. One natural question follows is that there exist dozens of SNRs near the solar system within 1 kiloparsec (kpc), as shown in Figure.\ref{fig:local_sources}, then, how is about the contribution from the other SNRs? In this work, we systematically study the contribution of all the local SNRs within 1 kpc around the solar system to the spectra of
  nuclei and electrons, the detailed parameters of these local SNRs are shown in Table \ref{tab:para_SNR}. The paper is organized as follows, Section 2 describes the model description, Section 3 presents the calculated results, and Section 4 gives the conclusion.

\begin{table*}
	\begin{center}
		\begin{tabular}{ccccccccc}
			\hline
			\hline
			$Number$ & SNR  & Other name & ~~~Distance~~~  & ~~~Radio index $\alpha_{r}$~~~ & ~~~Age~~~ & Pulsar &~~~References~~~ \\
			&G+long+lat & &[kpc] &  &[kyr] & &\\
			\hline
			1 &G194.3-13.1 &Geminga &0.33 & &345 &J0633+1746 &(1) \\
			2 &G65.3+5.7 & &$0.9\pm0.1$ &$0.58\pm0.07$ &$26\pm1$ & &(2-5)\\
			3 &G65.7+1.2 &DA 495 &$1.0\pm0.4$ &$0.45\pm0.1$  &$16.75\pm3.25$ &unkown &(6,7)\\
			4 &G74.0-8.5 &Cygnus Loop &$0.54_{-0.08}^{+0.10}$ &$0.4\pm0.06$  &$10\pm1$ &J1952+3252 &(5,8,9)\\
			5 &G114.3+0.3 & &0.7 &$0.49\pm0.25$ &$7.7\pm0.1$  & &(5,6,10,11)\\
			6 &G127.1+0.5 &R5 &$1\pm0.1$ &$0.43\pm0.1$ &$25\pm5$  & &(5,6,11-13)\\
			7 &G156.2+5.7 & &$0.8\pm0.5$ &$2.0_{-0.7}^{+1.1}$  &$10\pm1$ &B0450+55 &(5,6,14-17)\\
			8 &G160.9+2.6 &HB 9 &$0.8\pm0.4$ &$0.48\pm0.03$  &$5.5\pm1.5$ &B0458+46  &(5,6,11,18,19)\\
			9 &G203.0+12.0 &Monogem Ring &$0.288_{-0.027}^{+0.033}$ & &$86\pm1$ &B0656+14 &(20-21)\\
			10 &G263.9-3.3 &Vela(XYZ) &$0.295\pm0.075$ &variable  &$11.2\pm0.1$ &B0833-45 &(5,22-25)\\
			11 &G266.2-1.2 &Vela Jr &$0.75\pm0.01$ & &$3.5\pm0.8$ &J0855-4644 &(5,26-29)\\
			12 &G347.3-0.5 &SN393 &$1\pm0.3$ & &4.9 & &(5,30-31)\\
			\hline
		\end{tabular}\\
		{1\citep{1994A&A...281L..41S};2\citep{1996ApJ...458..257G};3\citep{2002A&A...388..355M};4\citep{2009A&A...503..827X};5\citep{2014BASI...42...47G};6\citep{2006A&A...457.1081K};7\citep{2008ApJ...687..516K};8\citep{2005AJ....129.2268B};9\citep{2006A&A...447..937S};10\citep{2004ApJ...616..247Y};11\citep{2014A&A...561A..55R};12\citep{1989A&A...219..303J};13\citep{2006A&A...451..251L};14\citep{1992A&A...256..214R};15\citep{2000bbxs.conf..567Y};16\citep{2007A&A...470..969X};17\citep{2009PASJ...61S.155K};18\citep{2003A&A...408..961R};19\citep{2007A&A...461.1013L};20\citep{1996ApJ...463..224P};21\citep{2003ApJ...593L..89B};22\citep{1999ApJ...515L..25C};23\citep{2001A&A...372..636A};24\citep{2001ApJ...561..930C};25\citep{2008ApJ...676.1064M};26\citep{1998Natur.396..141A};27\citep{1998Natur.396..142I};28\citep{2005MNRAS.356..969R};29\citep{2008ApJ...678L..35K};30\citep{2004ApJ...602..271L};31\citep{2009MNRAS.392..240M}.}
	\end{center}
	\caption{ Characteristic parameters of the nearby SNRs within 1 kpc}
	\label{tab:para_SNR}
\end{table*}

\section{Model Description}

%After entering the interstellar space, CRs undergo random walks within the galactic magnetic field by bouncing off the magnetic waves and magnetohydrodynamic turbulence.
After entering the interstellar space, CRs undergo the random walks within the galactic magnetic field by bouncing off the magnetic waves and magnetohydrodynamic turbulence. 
 The diffusive region, which is called the magnetic halo, is approximated as a flat cylinder with radius of $R=20$ kpc, equivalent to the galactic radius. The half-thickness $z_{h}$ is unknown, and is typically constrained by fitting the B/C ratio. The galactic disk, where both the CR sources and the interstellar gas are mainly spread across, is located in the middle of the magnetic halo. The width of the galactic disk is approximated to be invariant spatially and equals to $200$ pc. In addition to diffusion, CRs may also go through convection, diffusive-reacceleration, fragmentation, radioactive decay and other energy losses before arriving at the solar system. In fact, the process of convection is ignored in this work. The CR nuclei lose their energy principally via ionization, Coulomb scattering and adiabatic expansion. For electrons and positrons, their major energy loss mechanisms are the bremsstrahlung, synchrotron radiation and the inverse Compton scattering. This comprehensive process can be described by the propagation equation as,
\begin{equation}
\begin{array}{lcll}
\frac{\partial \psi(\vec{r},p,t)}{\partial t} &=& q(\vec{r}, p,t) + \vec{\nabla} \cdot
\left( D_{xx}\vec{\nabla}\psi - \vec{V_{c}}\psi \right)\\
&+& \frac{\partial}{\partial p}p^2D_{pp}\frac{\partial}{\partial p}\frac{1}{p^2}\psi
- \frac{\partial}{\partial p}\left[ \dot{p}\psi - \frac{p}{3}
\left( \vec{\nabla}\cdot \vec{V_c}\psi \right) \right]\\
&-& \frac{\psi}{\tau_f} - \frac{\psi}{\tau_r}
\end{array} ~,
\label{CRsPropagation}
\end{equation}
where $\rm q(\vec{r}, p,t)$ is the acceleration sources, $\rm \psi(\vec{r},p,t)$ the density of CR particles per unit momentum
$\rm p$ at position $\rm \vec{r}$,
$\rm \vec{V_c}$ the convection velocity,
$\rm \dot p\equiv dp/dt$ the momentum
loss rate, $\rm \tau_f$ and $\rm \tau _r$ the characteristic time scales for fragmentation and
radioactive decay, respectively, and
$\rm D_{xx}$ and $\rm D_{pp}$ the diffusion coefficients in the coordinate and the momentum space, respectively.

\subsection{Spatially-dependent propagation}

The SDP of CRs has received a lot of attention in recent years. It is first introduced as a Two Halo model (THM) \citep{2012ApJ...752L..13T} to explain the spectral hardening of both protons and helium above $200$ GeV \citep{2011Sci...332...69A}. Afterwards, it is further applied to the secondary and heavier components \citep{2015PhRvD..92h1301T, 2016PhRvD..94l3007F, 2016ApJ...819...54G, 2018ApJ...869..176L, 2020ChPhC..44h5102T, 2020arXiv200701768Y}, diffuse gamma-ray distribution \citep{2018PhRvD..97f3008G} and large-scale anisotropy \citep{2019JCAP...10..010L, 2019JCAP...12..007Q}. For a comprehensive introduction, one can refer to \citep{2016ApJ...819...54G} and \citep{2018ApJ...869..176L}.

In the SDP model, the whole diffusive halo is divided into two parts. The galactic disk and its surrounding area are called the inner halo (IH) region, in which the diffusion coefficient is spatially-dependent and relevant to the radial distribution of background CR sources. The extensive diffusive region outside the IH is named as the outer halo (OH) region, where the diffusion is regarded as only rigidity dependent. The spatially dependent diffusion coefficient $D_{xx}$ is thus parameterized as,
\begin{equation}
D_{xx}(r,z, {\cal R} )= D_{0}F(r,z)\beta^{\eta} \left(\dfrac{\cal R}{{\cal R}_{0}^{'}} \right)^{\delta(r,z)} ~,
\label{eq:diffusion}
\end{equation}
where $r$ and $z$ are cylindrical coordinate, ${\cal R}$ the particle's
rigidity, $\beta$ the particle's velocity in unit of light speed, and
$D_0$ and $\eta$ are constants. For the parameterization of $F(r,z)$ and
$\delta(r,z)$, one can refer to \citep{2020ChPhC..44h5102T}.
The total half-thickness of the propagation halo is $z_h$, and the
half-thickness of the inner halo is $\xi z_{h}$.
%For the parameterization of $F(r,z)$ and $\delta(r,z)$, one can refer to \cite{2020ChPhC..44h5102T}. The size of IH is represented by its half thickness $\xi z_{h}$, whereas the OH region's is $(1-\xi) z_{h}$.

In this work, we adopt the common diffusion-reacceleration (DR) model, with the diffusive-reacceleration coefficient $D_{pp}$ coupled to $D_{xx}$ by $D_{pp}D_{xx} = \dfrac{4p^{2}v_{A}^{2}}{3\delta(4-\delta^{2})(4-\delta)}$, where $v_A$ is the Alfv\'en velocity, $p$ is the momentum, and $\delta$ the rigidity dependence slope of the diffusion coefficient \citep{1994ApJ...431..705S}. The numerical package DRAGON is used to solve the SDP equation to obtain the distribution of the CR positrons, electrons and protons.
%During the propagation, the energetic CREs still suffer from the energy loss from synchrotron radiation and inverse Compton scattering \citep{1998ApJ...509..212S, 1998ApJ...493..694M}. A full-relativistic treatment of the inverse-Compton losses \citep{2010A&A...524A..51D} has been implemented in the DRAGON package \citep{2017JCAP...02..015E}.
Lower than tens of GeV, the CR fluxes are impacted by the solar modulation. The well-known force-field approximation \citep{1968ApJ...154.1011G, 1987A&A...184..119P} is applied to describe such an effect, with a modulation potential $\phi$ adjusted to fit the low energy data.

\subsection{Background sources}

The SNRs are regarded as the most likely sites for the acceleration of GCRs by default, in which the charge particles are accelerated to a power-law distribution through the diffusive shock acceleration. The distribution of the SNRs are approximated as axisymmetric, which is usually parameterized as,
\begin{equation}
f(r, z) = \left(\dfrac{r}{r_\odot} \right)^\alpha \exp \left[-\dfrac{\beta(r-r_\odot)}{r_\odot} \right] \exp \left(-\dfrac{|z|}{z_s} \right) ~,
\label{eq:radial_dis}
\end{equation}
where $r_\odot \equiv 8.5$ kpc represents the distance from the solar system to the galactic center. The parameters $\alpha$ and $\beta$ are taken as $1.69$ and $3.33$, respectively, in this work \citep{2015MNRAS.454.1517G}. The density distribution
of the SNRs decreases exponentially along the vertical height from the galactic plane, with $z_{s} = 200$ pc.

 %Past studies, for example Trotta R\citep{2011ApJ...729..106T}, Boschini \citep{2017ApJ...840..115B}; Yuan\citep{2020JCAP...11..027Y,}; Yuan\citep{2020JCAP...11..027Y}  have pointed out that the assumption of a single power-law injection spectrum could not reproduce the observations very well, especially in the diffusionreacceleration model. Furthermore, this low-energy break has also been indirectly supported by the observations of nearby molecular clouds, see A.Neronov\citep{2012PhRvL.108e1105N}. Therefore we introduce a break at low energy of the injection spectrum.
%Past studies, for example \citep{2011ApJ...729..106T,2017ApJ...840..115B,2020JCAP...11..027Y} have pointed out that the assumption of a single power-law injection spectrum could not reproduce the observations very well, especially in the diffusionreacceleration model. Furthermore, this low-energy break has also been indirectly supported by the observations of nearby molecular clouds, see A.Neronov\citep{2012PhRvL.108e1105N}. Therefore we introduce a break at low energy of the injection spectrum.

Past studies, for example, \citep{2011ApJ...729..106T,2017ApJ...840..115B,2020JCAP...11..027Y}, have pointed out that the assumption of a single power-law injection spectrum could not reproduce the observations very well, especially in the diffusion-reacceleration model. Furthermore, this low-energy break has also been indirectly supported by the observations of nearby molecular clouds, see A. Neronov et al. \citep{2012PhRvL.108e1105N}. Therefore, we introduce a break at low energy of the injection spectrum,
%The injection spectra of nuclei and electron are assumed to be exponentially cutoff broken power-law function of particle rigidity ${\cal R}$, i.e.
\begin{equation}\label{eq:spectrum_CRE}
q({\cal R}) = q_{0} \left\{
\begin{array}{lll}
	\left(\dfrac{{\cal R}}{{\cal R}_{\rm br1}} \right)^{\nu_{1}}, ~{{\cal R} \leq {\cal R}_{\rm br1}}\\
	\\
	\left(\dfrac{{\cal R}}{{\cal R}_{\rm br1}} \right)^{\nu_{2}} \exp\left[-\dfrac{\cal R}{{\cal R}_{\rm c}} \right],    ~{ {\cal R} > {\cal R}_{\rm br1}}
\end{array} ~,
\right.
\end{equation}
where $q_{0}$ is the normalization factor, $\nu_{1,2}$ the spectral
incides, ${\cal R}_{\rm br1}$ break rigidities, and ${\cal R}_{\rm c}$
the cutoff rigidity.
%, and ${\cal R}_0$ is a reference rigidity.
%\begin{equation}\label{eq:spectrum_CRE}
%\mathcal{Q}({\cal R}) = \mathcal{Q}_{0} \left\{
%\begin{array}{lll}
%\left(\dfrac{{\cal R}_{\rm br1} }{{\cal R}_{0}} \right)^{\nu_{2}}
%\left(\dfrac{{\cal R}}{{\cal R}_{\rm br1}} \right)^{\nu_{1}},  & {{\cal R} \leqslant {\cal R}_{\rm br1}} \\
%\\
%\left(\dfrac{{\cal R}}{{\cal R}_{0}} \right)^{\nu_{2}},  & {\cal R}_{\rm br1} < {\cal R} \leqslant {\cal R}_{\rm br2} \\
%\\
%\left(\dfrac{{\cal R}_{\rm br2} }{{\cal R}_{0}} \right)^{\nu_{2}} \left(\dfrac{{\cal R}}{{\cal R}_{\rm br2}} \right)^{\nu_{3}},  & {\cal R} > {\cal R}_{\rm br2}
%\end{array}
%\right.
%\end{equation}
%%\textcolor {red}
%{where $\mathcal{Q}_0$, $\nu_{1/2/3}$, $R_{0}$ and $R_{\rm br1/2}$ are the normalization, power index, reference rigidity and broken rigidity respectively.}

\subsection{Local SNRs}

At $\sim$ TeV energies, CR electrons are found to originate from local sources within $\sim1$ kpc around
the solar system \citep{2018SCPMA..61j1002Y}. In this small region, the hypothesis of continuous distribution may not be valid any more. Studies show that the discrete effect of the nearby CR sources could induce large fluctuations, especially at high energies \citep{2011JCAP...02..031M,2012A&A...544A..92B,2017ApJ...836..172F}. The contribution of the nearby sources to the CR nuclei and electrons have been studied in the past works (see e.g., \citep{2012APh....39....2S, 2014JCAP...04..006D,2017PhRvD..96b3006L,2018ApJ...854...57F}). In this work, we assume that a nearby SNR accounts for the excess of electrons (protons) above $\sim 100$ (or 200) GeV, respectively.
The propagation of CRs injected instantaneously from a point source
is described by a time-dependent propagation equation
\citep{1995PhRvD..52.3265A}. The injection rate as a function of time
and rigidity is assumed to be,
\begin{equation}
Q^{\rm snr}({\cal R},t)=Q^{\rm snr}_{0}(t) \left(\dfrac{\cal R}{{\cal R}_0}
\right)^{-\gamma} \exp \left[-\dfrac{\cal R}{{\cal R}^{\rm e^{-},p}_{\rm c}}
\right] ~,
\label{eq:nearby}
\end{equation}
where ${\cal R}^{\rm e^{-},p}_{\rm c}$ is the cutoff rigidity of its accelerated CRs. %A continuous injection process of electron and positron pairs with injection rate proportional to the spindown power of the pulsar is assumed, i.e.,
%\begin{equation}
%Q_0^{\rm psr} (t) = \dfrac{q_0^{\rm psr}}{(1+t/\tau_0)^2} ~,
%\end{equation}
%where $\tau_0$ is a characteristic time scale of the decay of the spindown \citep{2010ApJ...710..958K, 2013PhRvD..88b3001Y}.

The local SNRs may accelerate primary nuclei and electrons during their early evolution stage. This contribution of primary electrons from local sources may be necessary given the different spectral behaviors of the positrons and the electrons \citep{2021JCAP...05..012Z}. The injection process of the SNRs is
approximated as burst-like. The source injection rate is assumed as the following,
\begin{equation}
Q_{0}^{\rm snr}(t) = q_{0}^{\rm snr} \delta(t-t_0) ~, \\
\end{equation}
where $t_0$ is the time of the supernova explosion. The propagated spectrum from the local SNRs is thus a convolution of the
Green's function and the time-dependent injection rate $Q_0(t)$
\citep{1995PhRvD..52.3265A},
\begin{equation}
\varphi(\vec{r}, {\cal R}, t) = \int_{t_i}^{t} G(\vec{r}-\vec{r}^\prime, t-t^\prime, {\cal R}) Q_0(t^\prime) d t^\prime .
\end{equation}

\section{Results}
\begin{figure}[!htb]
	\centering
	\includegraphics[height=6.5cm, angle=0]{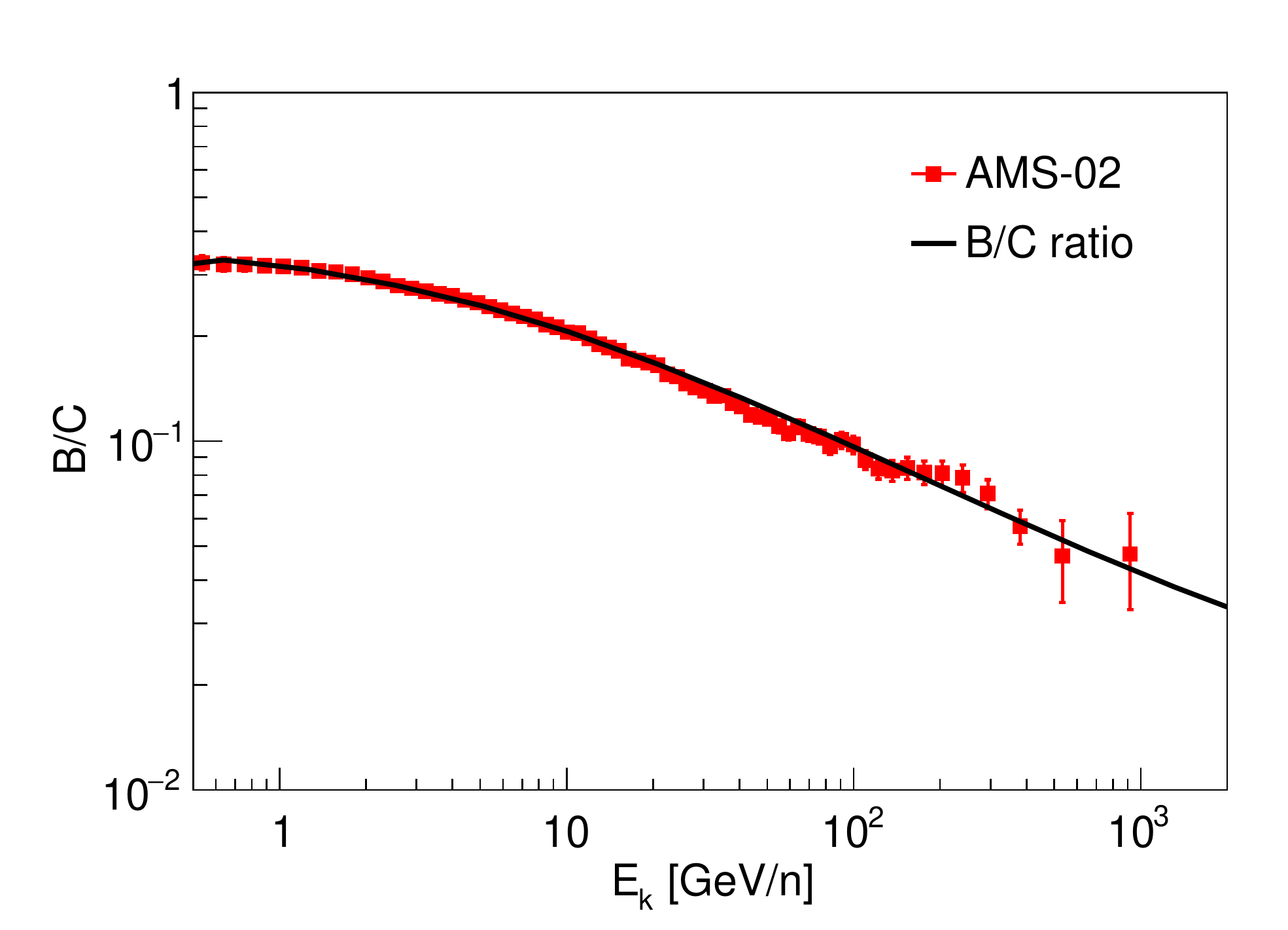}
	\caption{Model prediction of the B/C ratio compared with the AMS-02 measurement \citep{2015PhRvL.114q1103A,2017PhRvL.119y1101A}.
	}
	\label{fig:BC}
\end{figure}

In this work, both the propagation and the injection parameters are fitted manually. By fitting the B/C ratio, the diffusion coefficient parameters are $D_0 = 4.87\times 10^{28}$
cm$^2$~s$^{-1}$, $\delta_0 = 0.55$, $N_m = 0.6$, $n = 4$, $\xi=0.1$,
and $\eta=0.05$. The half thickness of the propagation halo is $z_h=5$ kpc,
and the Alfv\'enic velocity is $v_A=6$ km~s$^{-1}$.
 The comparison of the B/C ratio between the model prediction and the observation data is given in Figure
\ref{fig:BC}, which indicates that the relevant parameters are reasonably matched. In order to determine which nearby SNRs are the appropriate candidates on the basis of their contributions to the proton and electron spectra, respectively, we carry out more detailed analyses in the following.

\subsection{Spectra of all the local SNRs}
\begin{figure*}[!htb]
	\centering
	\includegraphics[height=7.5cm, angle=0]{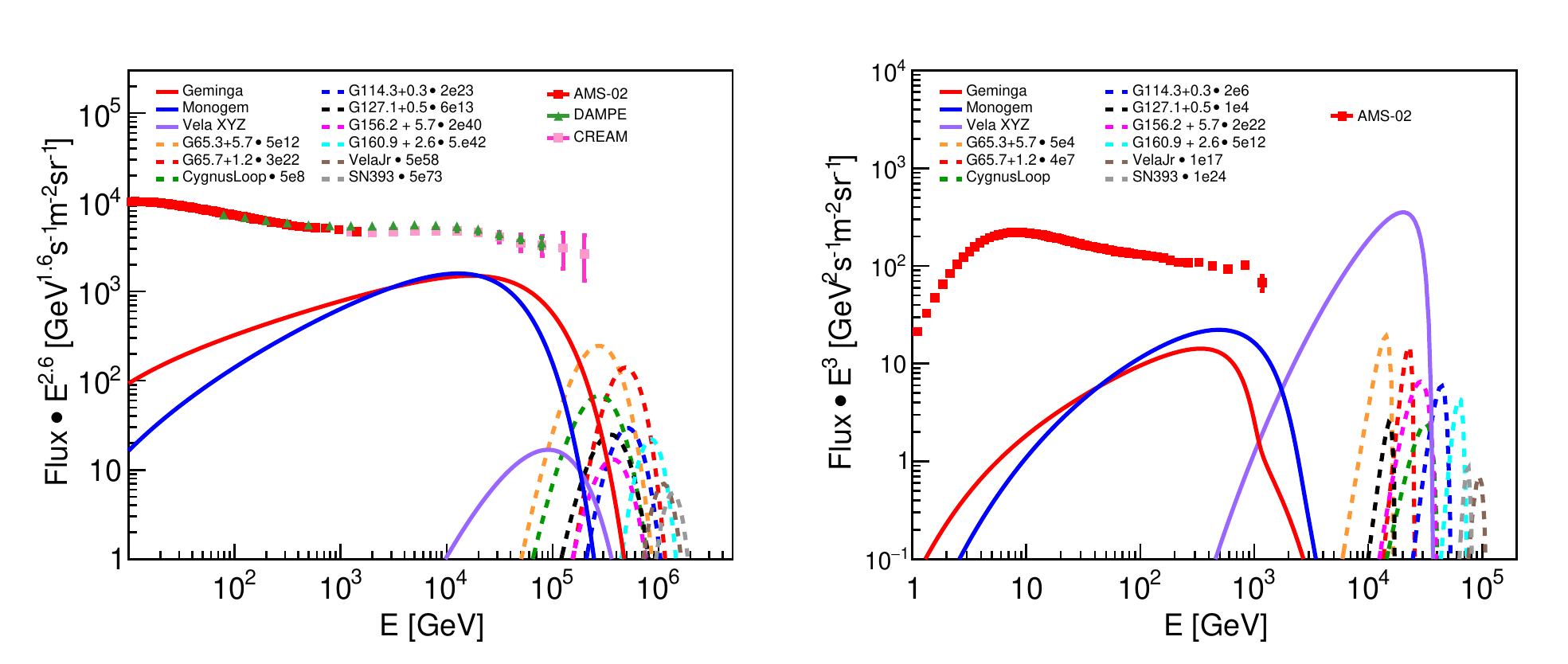}
	\caption{The calculated energy spectra of protons (left) and electrons (right) from the different local SNRs. The red, blue and purple solid lines represent the energy spectra of the Geminga, the Monogem, and the Vela SNRs, respectively. The dashed lines with different colors represent the energy spectra of other local SNRs, respectively. For comparing the results conveniently, the fluxes of some local SNRs are multiplied by a scale factor. The red-square, green-triangle and magenta-square data points are from the AMS-02, the DAMPE, and the CREAM experiments, respectively \citep{2015PhRvL.114q1103A,2015PhRvL.115u1101A,2017ApJ...839....5Y,2019SciA....5.3793A}.
	}
	\label{fig:AllSource}
\end{figure*}

Figure \ref{fig:AllSource} presents the calculated proton and electron spectra for all the SNRs as listed in Table \ref{tab:para_SNR}. Here, it can be seen that only the Geminga, the Monogem, and the Vela SNRs have significant contributions to the proton and electron spectra, respectively.
Whereas most others have very limited contribution to the spectra. To appear in the same figure, we
multiply a scale factor in the fluxes for the faint SNRs to show properly, as shown by the dashed line in Figure \ref{fig:AllSource}. The detailed parameters of the local sources are listed in Table \ref{tab:snr_inj}. In the following subsection, we give the detailed studies on the contribution to the spectra and anisotropy by the Geminga, the Monogem, and the Vela SNRs. By doing so, we can expect to pinpoint the optimal local SNR.

\subsection{Spectra and anisotropy of the Geminga SNR}
\begin{figure*}[!htb]
	\centering
	\includegraphics[height=13cm, angle=0]{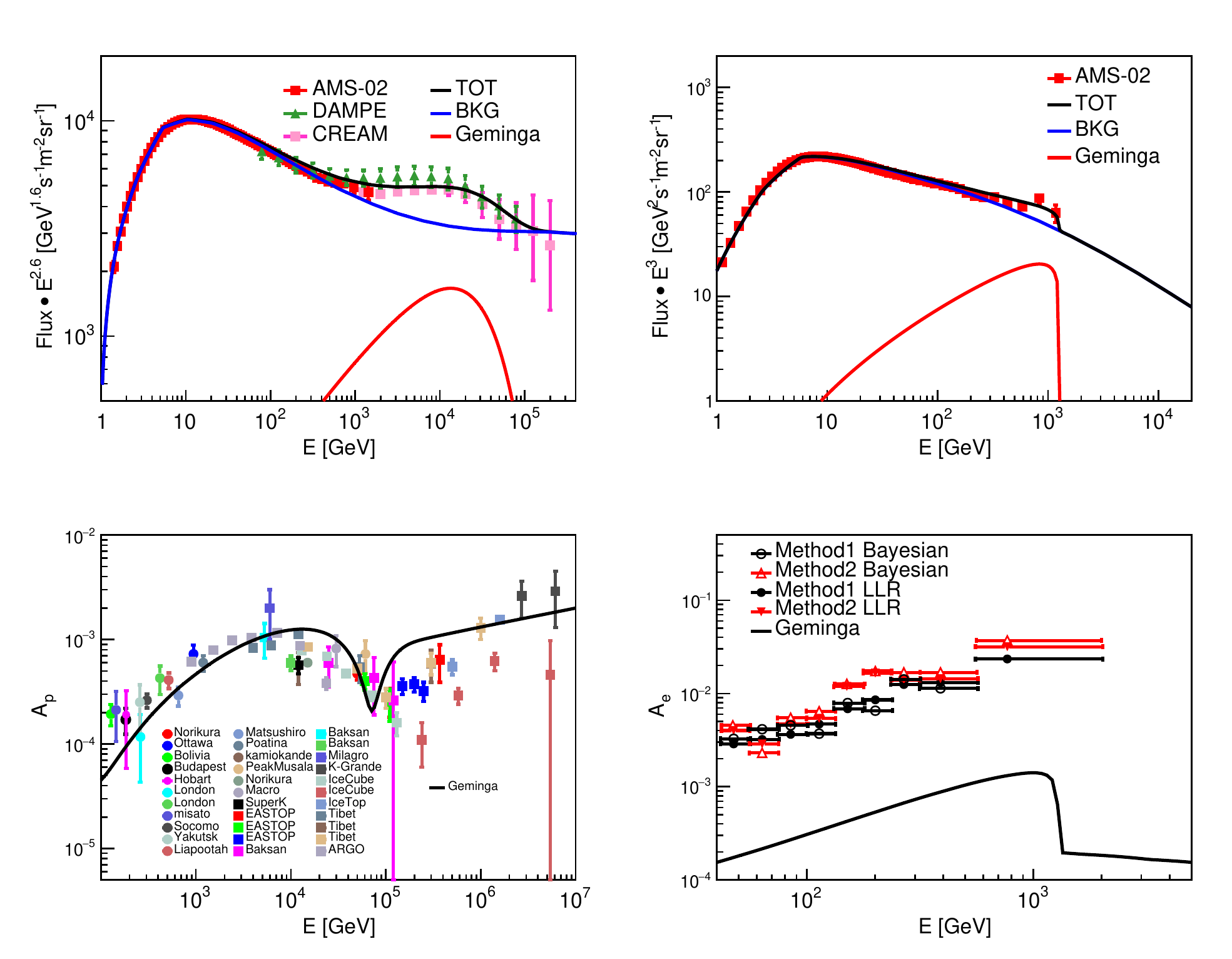}
	\caption{The calculated energy spectra (top) and anisotropy (bottom) for protons (left) and primary electrons (right) from the Geminga SNR, respectively. For the energy spectra, the blue lines represent the fluxes from the background sources, the red lines the fluxes from the local Geminga SNR, and the black solid lines their corresponding sum. For the anisotropies, the black solid lines are the model expected results.
	}
	\label{fig:Geminga}
\end{figure*}

Figure \ref{fig:Geminga} illustrates the calculated energy spectra and anisotropy of protons and primary electrons, respectively. The blue solid lines are the fluxes from the background sources. Taking into account the contributions of the Geminga SNR as shown in Figure \ref{fig:AllSource}, the model predictions of the proton and primary electron spectra are in good agreement with the experimental data, as shown in the upper panel of Figure \ref{fig:Geminga}. Table \ref{tab:snr_inj} shows the relevant background and local sources parameters. In fact, the anisotropy is a more effective factor to determine the dominant local SNRs from their direction, age and distance.
The bottom panel of Figure \ref{fig:Geminga} shows the model calculated anisotropy for CRs and electrons, respectively.
It is obvious that the model calculated anisotropy is consistent with that of the observations of CRs. And the electron anisotropy is roughly at the same level and far below the Fermi-LAT limit.
Therefore, the results support our previous conclusion of the co-evolved spectrum and anisotropy for the Geminga SNR \citep{2019JCAP...10..010L,2019JCAP...12..007Q,2021JCAP...05..012Z}.

\begin{table*}
	\begin{center}
		\begin{tabular}{|c|cccccc|}
			\hline
			Bckgroud  & Normalization$^\dagger$ & ~~~$\nu1$~~~  & ~~~$\mathcal R_{br1}$~~~ &  ~~~$\nu2$~~~  & ~~~$\mathcal R_{c}$~~~~~ & \\
			& $[({\rm m}^2\cdot {\rm sr}\cdot {\rm s}\cdot {\rm GeV})^{-1}]$ & & [GV] & & [GV] &\\
			\hline
			P & $4.62\times 10^{-2}$   & 2  & $6$  & 2.41 &$7\times 10^{6}$  & \\
			Electron & $2.80\times 10^{-1}$   & 1.50    &  5.10  & 2.80 &300  & \\
			\hline
			SNR  & $r_{snr}$ &$t_{snr}$ &$\gamma_{snr}$ &$\mathcal R_{c}$ &$q_{0}^{prot}$ &$q_{0}^{elec}$\\
			&[kpc] & [yrs]  & &[TV] &$[\rm GeV^{-1}]$ &$[\rm GeV^{-1}]$\\
			\hline
			Geminga &0.33 &$3.45\times 10^{5}$  &2.13  &55 &$2.50\times10^{52}$ &$1.20\times10^{50}$\\
			Monogem &0.29 &$8.60\times 10^{4}$ &2.13 &27 &$1.20\times10^{52}$ &$2.10\times10^{49}$\\
			Others  &... &... &$2\alpha_{r} + 1$ &55 &$2.50\times10^{52}$ &$1.20\times10^{50}$\\
			\hline
		\end{tabular}\\
		$^\dagger${The proton and electron normalization are set at kinetic energy per nucleon $E_{k} = 100$ GeV/n and $10$ GeV/n, respectively.}
	\end{center}
	\caption{The background and local SNRs injected parameters for protons and electrons. For the age and distance parameters of the local sources, refer to Table \ref{tab:para_SNR}}
	\label{tab:snr_inj}
\end{table*}

\subsection{Spectra and anisotropy of the Monogem SNR}
\begin{figure*}[!htb]
	\centering
	\includegraphics[height=13cm, angle=0]{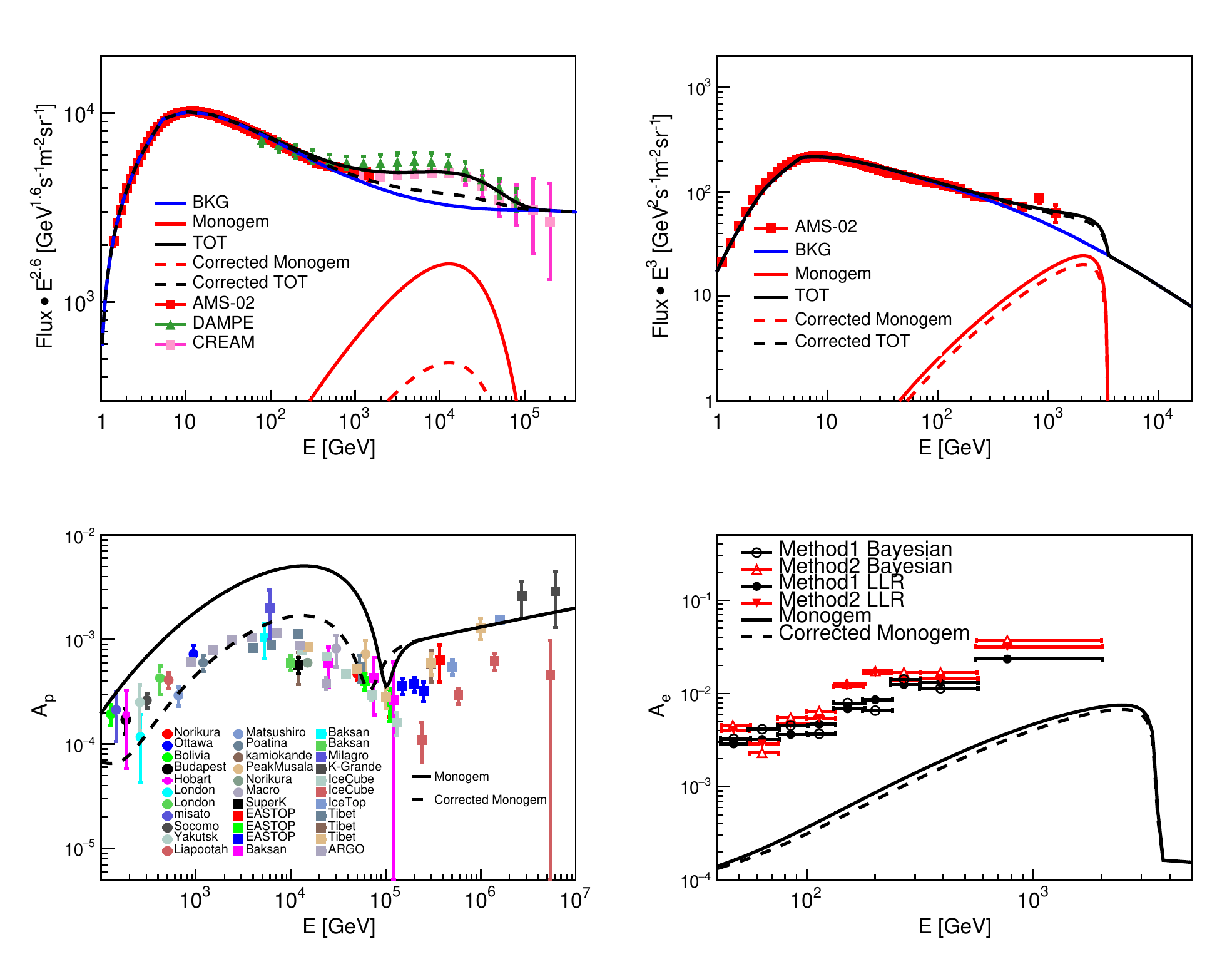}
    \caption{The calculated energy spectra (top) and anisotropy (bottom) for protons (left) and primary electrons (right) from the Monogem SNR, respectively. The implication of different color solid lines is the same as that in Figure \ref{fig:Geminga} and the different colored dashed lines display the corrected injection energy in accordance with the spin-down energy of the Geminga and the Monogem pulsars.
    }
	\label{fig:Monogem}
\end{figure*}

Similar to the results of the Geminga SNR, the model calculations of the Monogem SNR are
presented in Figure \ref{fig:Monogem}. The proton spectrum is also consistent with the
observations. However, the calculated anisotropy greatly overthrows the observations.
Obviously, there exists tension between the spectrum and the anisotropy.
In addition, the cut-off energy of the electron spectrum is also beyond the TeV energy.

It should be noticed that the spin-down energy of the Monogem pulsar is lower than that of the Geminga pulsar with
the value of $1.8\times10^{48}/1.25\times10^{49} erg$. It is possible that the injection power of the Monogem SNR is
much lower than that of the Geminga SNR. In fact, the $\gamma$-ray emission at TeV energy is much lower for the
Monogem compared with that of the Geminga \citep{2017Sci...358..911A}.
In Figure \ref{fig:Monogem}, the dashed line shows the corrected
injection energy resulted from the spin-down energy. Under these circumstances, the contribution from the Monogem SNR can be ignored.

Another reason is the propagation effect in the source region. It is well known that
the propagation coefficient in the source region is much lower \citep{2017Sci...358..911A}. This means that
the CRs spend more time to disentangle from the source. The age of Monogem is at the
level of 86 kyrs and only a limited part of the CRs from Monogem arrives at our solar system.

On the other hand, the regular magnetic field could also regulate the CR flux
from the local source. The observations \citep{2014Sci...343..988S,2009Sci...326..959M,2016ApJ...818L..18Z} have revealed that the phase of
anisotropy less than 100 TeV is consistent with the effect of the local regular magnetic
field. Ahlers \citep{2014PhRvL.112b1101A} has pointed out that the local regular magnetic field
and the corresponding anisotropic diffusion have to be considered when accounting
for the evolution relation of the amplitude and phase of the anisotropy with energy. When
the regular magnetic field is taken into account, the CR flux is projected into
the regular magnetic field line and propagates along the magnetic field. If
the sources are far away from the regular magnetic field, their flux to the
solar system could be suppressed by the regular magnetic field. This would
be investigated in detail in the coming work.

In a word, the contribution from the Monogem SNR is possibly limited. Other information is still required to check our model further on this point.

\subsection{Spectra and anisotropy of the Vela SNR}
\begin{figure*}[!htb]
	\centering
	\includegraphics[height=6.6cm, angle=0]{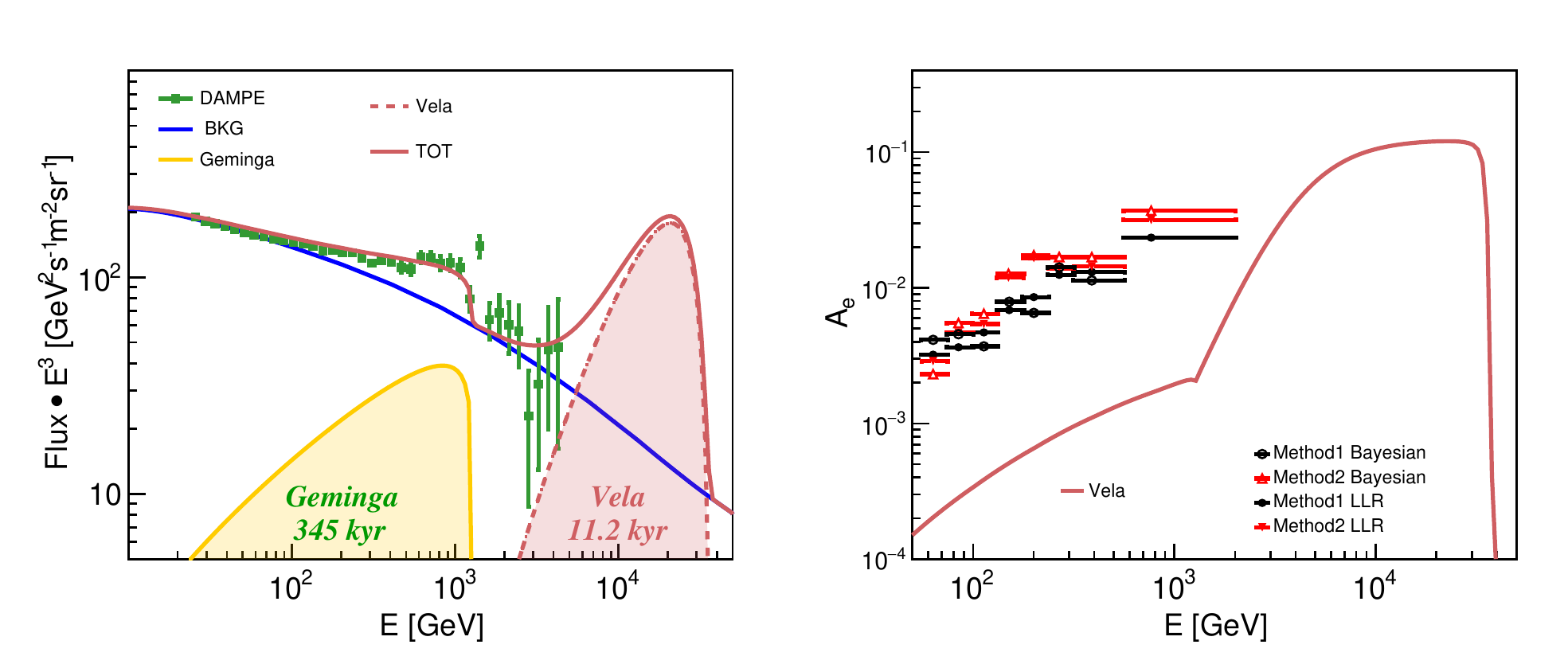}
	\caption{Left: The calculated total positron and electron spectra of the Vela SNR. The blue solid line is the total background flux. The yellow solid line with shadow indicates the Geminga SNR flux. The light-red dashed line with shadow represents the flux from the Vela SNR for different injected energies. The corresponding solid line is the sum of the background, the Geminga, and the Vela fluxes. The green squares are the measurements from the DAMPE experiment \citep{2017Natur.552...63D}. Right: Anisotropies of electrons with the Geminga and the Vela SNRs under the SDP model. %The implication of solid lines with different colors are the same as the left panel.
	Both the black and red markers are the upper limits set by the {\em Fermi}-LAT experiment \citep{2017PhRvL.118i1103A}.
	}
	\label{fig:vela}
\end{figure*}

\begin{figure*}[!htb]
	\centering
	\includegraphics[height=13cm, angle=0]{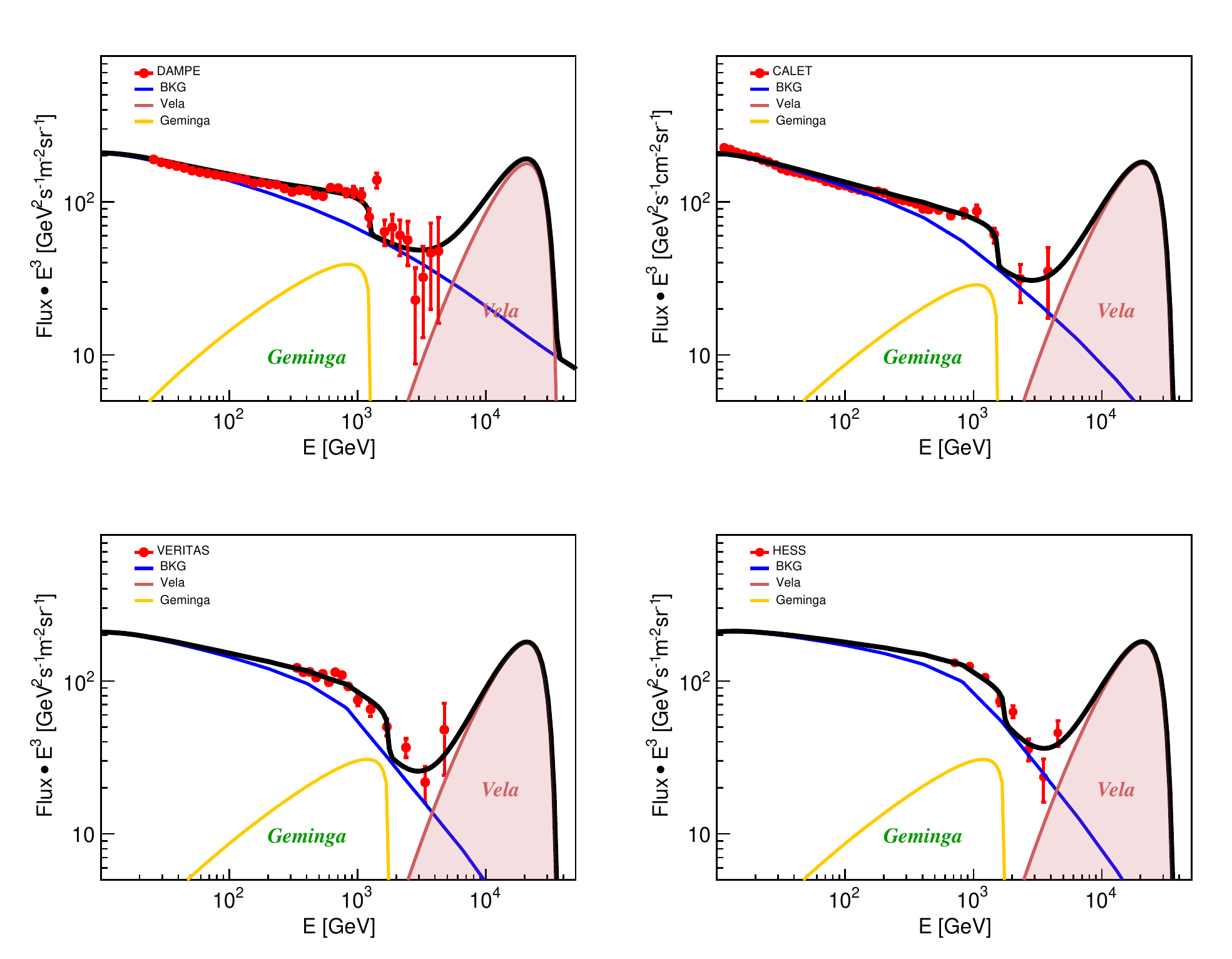}
	\caption{Our model expectations of the total electron and positron spectra compared with the experiments of DAMPE (left-top), CALET (right-top), VERITAS (left-bottom) and HESS (right-bottom), respectively. The blue solid lines represent the fluxes from the background sources, the yellow lines are the fluxes from the Geminga SNR, the red lines with shadow are the fluxes from the Vela SNR, and the black lines indicate the corresponding sum of the above components.
	}	
	\label{fig:velafit}
\end{figure*}

 The Vela SNR is a very young and nearest celestial body. It can provide very important clue
 to check our model in terms of the CR spectrum and anisotropy. Figure \ref{fig:vela} shows the expected spectrum and anisotropy for electrons. The related parameters are shown in Table \ref{tab:para_elec}. It is obvious that there exists a bump structure in the spectrum and sharp rising
 in the anisotropy above the TeV energy. In fact, the bump structure in the spectrum has been hinted
 by the HESS, the VERTITAS, the DAMPE, and the CALET observations \citep{2017Natur.552...63D,2021Investigating,2018PhRvD..98f2004A,2008PhRvL.101z1104A}, as shown in Figure \ref{fig:velafit}.
 In addition, the electron anisotropy reveals an remarkable hoist, which reaches $10\%$ at several TeV.
 What makes the fact more convincible is that those abnormal phenomena can be tested soon.
 Firstly, the space-borne experiments DAMPE and HERD will be observing the new spectral
 structure in the short run. Simultaneously, the anisotropy is overwhelmingly larger for the electrons than for the CRs,
 which makes it easily observed by the ground-based experiments HAWC and LHAASO-WCDA.

\begin{figure*}[!htb]
	\centering
	\includegraphics[height=13cm, angle=0]{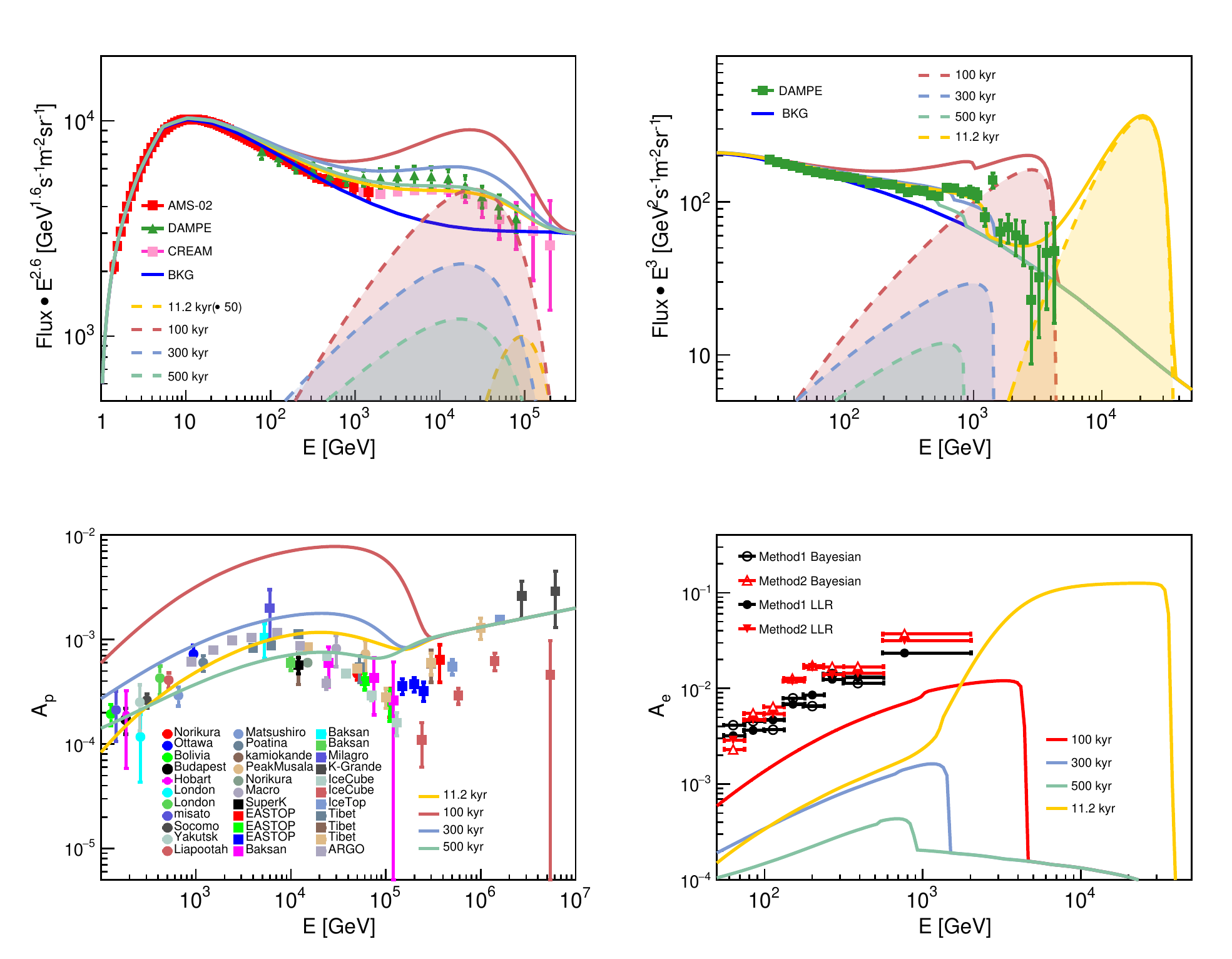}
	\caption{The calculated energy spectra (top) and anisotropy (bottom) for protons (left) and primary electrons (right) from the Vela SNR, respectively. For the energy spectra, the blue lines are the fluxes from the background sources, the different colored dashed lines are the fluxes from the Vela SNR for various ages, and the different solid lines are their corresponding sum including the Geminga SNR flux. Note that the Geminga SNR spectrum is not shown here. For the anisotropies, the different colored solid lines are the model expected results.
	}
	\label{fig:vela2}
\end{figure*}

 Considering the importance of the Vela SNR, it is necessary to perform further study such as on the effect of
 different ages. Figure \ref{fig:vela2} shows the spectra and anisotropy at 10, 100, 300, and 500 kyrs for
 protons and electrons, respectively. It is obvious that the flux reaches its maximum at 100 kyrs and then decreases smoothly. When the source age becomes as old as 500 kyrs, the contribution can be ignored.
 In a word, the Vela SNR plays the similar role of Monogem at $\sim$ 100 kyrs, and that of Geminga at $\sim$ 300 kyrs, and disappears above 500 kyrs.

\begin{table*}
	\begin{center}
		\begin{tabular}{|c|cccccc|}
			\hline
			Bckgroud  & Normalization$^\dagger$ &$\nu1$ &$\mathcal R_{br1}$ &$\nu2$ &$R_{c}$ &\\
			%			\hline
		    (DAMPE)& $[({\rm m}^2\cdot {\rm sr}\cdot {\rm s}\cdot {\rm GeV})^{-1}]$ & & [GV] & &[GV] &\\
			\hline
			Electron & $2.55\times 10^{-1}$   & 1.60    &  5.10  & 2.72 & 1200 &  \\
%			CALET & $2.80\times 10^{-1}$   & 1.60    &  5.10  & 2.76 & 600 &3.00  \\
%			VERITAS & $2.80\times 10^{-1}$   & 1.60    &  5.10  & 2.70 & 750 &3.40  \\
%			HESS & $2.80\times 10^{-1}$   & 1.60    &  5.10  & 2.62 & 1000 &3.35  \\
			\hline
			SNR  & $r_{snr}$ &$t_{snr}$ &$\gamma_{snr}$ &$\mathcal R_{c}$ &$q_{0}^{elec}$ &\\
			&[kpc] & [yrs]  & &[TV] &$[\rm GeV^{-1}]$ &\\
			\hline
			Geminga &0.33 &$3.45\times 10^{5}$  &2.13  &55 &$2.30\times10^{50}$ &\\
			Vela &0.30 &$1.12\times 10^{4}$ &2.10 &55 &$5.00\times10^{49}$ &\\
			\hline
		\end{tabular}\\
		$^\dagger${The normalization is set at kinetic energy per nucleon $E_{k} = 10$ GeV/n.}
	\end{center}
	\caption{Electron background and local SNRs injected parameters about the DAMPE experiment}
	\label{tab:para_elec}
\end{table*}

\section{Conclusion}
{

With the operation of new generations of space-borne and ground-based CR experiments, the measurements of CRs have stepped into a precise era. 
More and more fine structures have been unveiled in the CR energy spectra and anisotropy from hundreds of GeV to hundreds of TeV. To reproduce these observational phenomena, the local sources have to be involved in the model calculation. 

In this work, we systematically study the contribution of all the local SNRs within 1 kpc around the solar system to the spectra and anisotropy of CR nuclei and electrons. We demonstrate that three local SNRs, i.e. Geminga, Monogem, and Vela, could have important contribution to both proton and electron spectra, as inferred from their distances and ages, assuming a common injection energy. However, the expected anisotropy from Monogem is obviously inconsistent with the observations. One of possibilities is its injection power is lower, considering that the spin-down energy of the Monogem pulsar is lower about an order of magnitude than that of the Geminga pulsar. When the total injection energy lowered by one order of magnitude, the influence from Monogem can be safely neglected. That leaves the Gminga SNR as the dominant source responsible for the observational anomalies of the nuclear spectra and the anisotropy.

Furthermore, we expected that there is a new bump in electron spectrum above several TeV, which could stem from the young Vela SNR. Simultaneously, the magnitude of anisotropy can reach as high as 10\% at the corresponding energy. Even the injected energy reuduces by an order of magnitude, the excesses from Vela SNR in the spectrum and the anisotropy are still observable. We hope such excesses are going to be checked soon by the space-borne DMAPE and HERD and the ground-based HAWC and LHAASO experiments.

One has to be noted that the CR diffusion adopted in the above calculation is approximated as isotropic, in which corresponds to the case of an ideal turbulent magnetic field in plasma. In fact, it is well-known that there exists a large-scale regular magnetic field in Galaxy. When the regular component is taken into account, the diffusion coefficient is converted to a tensor, i.e. the diffusion coefficient parallel and perpendicular to the regular magnetic field have to be considered respectively\citep{2017JCAP...10..019C,2020ApJ...892....6L}. In the coming work, we would like to further study the influence of the regular magnetic field on the energy spectra and anisotropies.

\section*{Acknowledgements}
This work is supported by the National Key R\&D Program of China under Grant No. 2018YFA0404202, and the National Natural Science Foundation of China (Nos. 11635011, 11875264, U1831129, U1831208, and U2031110).

\textbf{\textit{Note added in proof}}. While this paper was ready to submit, another similar work is noticed to have appeared in https://arxiv.org/\citep{tang2021explanation}.

\bibliographystyle{apj}
\bibliography{ref1}

\begin{thebibliography}{}
\expandafter\ifx\csname natexlab\endcsname\relax\def\natexlab#1{#1}\fi

\bibitem[{{Aartsen} {et~al.}(2013){Aartsen}, {Abbasi}, {Abdou}, {Ackermann},
  {Adams}, {Aguilar}, {Ahlers}, {Altmann}, {Andeen}, {Auffenberg}, \&
  et~al.}]{2013ApJ...765...55A}
{Aartsen}, M.~G., {Abbasi}, R., {Abdou}, Y., {et~al.} 2013, \apj, 765, 55

\bibitem[{{Aartsen} {et~al.}(2016){Aartsen}, {Abraham}, {Ackermann}, {Adams},
  {Aguilar}, {Ahlers}, {Ahrens}, {Altmann}, {Anderson}, {Ansseau}, {Anton},
  {Archinger}, {Arguelles}, {Arlen}, {Auffenberg}, {Bai}, {Barwick}, {Baum},
  {Bay}, {Beatty}, {Becker Tjus}, {Becker}, {Beiser}, {BenZvi}, {Berghaus},
  {Berley}, {Bernardini}, {Bernhard}, {Besson}, {Binder}, {Bindig}, {Bissok},
  {Blaufuss}, {Blumenthal}, {Boersma}, {Bohm}, {B{\"o}rner}, {Bos}, {Bose},
  {B{\"o}ser}, {Botner}, {Braun}, {Brayeur}, {Bretz}, {Buzinsky}, {Casey},
  {Casier}, {Cheung}, {Chirkin}, {Christov}, {Clark}, {Classen}, {Coenders},
  {Collin}, {Conrad}, {Cowen}, {Cruz Silva}, {Daughhetee}, {Davis}, {Day}, {de
  Andr{\'e}}, {De Clercq}, {del Pino Rosendo}, {Dembinski}, {De Ridder},
  {Desiati}, {de Vries}, {de Wasseige}, {de With}, {DeYoung},
  {D{\'\i}az-V{\'e}lez}, {di Lorenzo}, {Dujmovic}, {Dumm}, {Dunkman},
  {Eberhardt}, {Ehrhardt}, {Eichmann}, {Euler}, {Evenson}, {Fahey}, {Fazely},
  {Feintzeig}, {Felde}, {Filimonov}, {Finley}, {Flis}, {F{\"o}sig}, {Fuchs},
  {Gaisser}, {Gaior}, {Gallagher}, {Gerhardt}, {Ghorbani}, {Gier}, {Gladstone},
  {Glagla}, {Gl{\"u}senkamp}, {Goldschmidt}, {Golup}, {Gonzalez}, {G{\'o}ra},
  {Grant}, {Griffith}, {Ha}, {Haack}, {Haj Ismail}, {Hallgren}, {Halzen},
  {Hansen}, {Hansmann}, {Hansmann}, {Hanson}, {Hebecker}, {Heereman},
  {Helbing}, {Hellauer}, {Hickford}, {Hignight}, {Hill}, {Hoffman}, {Hoffmann},
  {Holzapfel}, {Homeier}, {Hoshina}, {Huang}, {Huber}, {Huelsnitz}, {Hulth},
  {Hultqvist}, {In}, {Ishihara}, {Jacobi}, {Japaridze}, {Jeong}, {Jero},
  {Jones}, {Jurkovic}, {Kappes}, {Karg}, {Karle}, {Katz}, {Kauer}, {Keivani},
  {Kelley}, {Kemp}, {Kheirandish}, {Kim}, {Kintscher}, {Kiryluk}, {Klein},
  {Kohnen}, {Koirala}, {Kolanoski}, {Konietz}, {K{\"o}pke}, {Kopper}, {Kopper},
  {Koskinen}, {Kowalski}, {Krings}, {Kroll}, {Kroll}, {Kr{\"u}ckl}, {Kunnen},
  {Kunwar}, {Kurahashi}, {Kuwabara}, {Labare}, {Lanfranchi}, {Larson},
  {Lennarz}, {Lesiak-Bzdak}, {Leuermann}, {Leuner}, {Lu}, {L{\"u}nemann},
  {Madsen}, {Maggi}, {Mahn}, {Mandelartz}, {Maruyama}, {Mase}, {Matis},
  {Maunu}, {McNally}, {Meagher}, {Medici}, {Meier}, {Meli}, {Menne}, {Merino},
  {Meures}, {Miarecki}, {Middell}, {Mohrmann}, {Montaruli}, {Morse},
  {Nahnhauer}, {Naumann}, {Neer}, {Niederhausen}, {Nowicki}, {Nygren},
  {Obertacke Pollmann}, {Olivas}, {Omairat}, {O'Murchadha}, {Palczewski},
  {Pandya}, {Pankova}, {Paul}, {Pepper}, {P{\'e}rez de los Heros}, {Pfendner},
  {Pieloth}, {Pinat}, {Posselt}, {Price}, {Przybylski}, {Quinnan}, {Raab},
  {R{\"a}del}, {Rameez}, {Rawlins}, {Reimann}, {Relich}, {Resconi}, {Rhode},
  {Richman}, {Richter}, {Riedel}, {Robertson}, {Rongen}, {Rott}, {Ruhe},
  {Ryckbosch}, {Sabbatini}, {Sander}, {Sandrock}, {Sandroos}, {Sarkar},
  {Schatto}, {Schimp}, {Schlunder}, {Schmidt}, {Schoenen}, {Sch{\"o}neberg},
  {Sch{\"o}nwald}, {Schumacher}, {Seckel}, {Seunarine}, {Soldin}, {Song},
  {Spiczak}, {Spiering}, {Stahlberg}, {Stamatikos}, {Stanev}, {Stasik},
  {Steuer}, {Stezelberger}, {Stokstad}, {St{\"o}ssl}, {Str{\"o}m},
  {Strotjohann}, {Sullivan}, {Sutherland}, {Taavola}, {Taboada}, {Tatar},
  {Ter-Antonyan}, {Terliuk}, {Te{\v{s}}i{\'c}}, {Tilav}, {Toale}, {Tobin},
  {Toscano}, {Tosi}, {Tselengidou}, {Turcati}, {Unger}, {Usner}, {Vallecorsa},
  {Vandenbroucke}, {van Eijndhoven}, {Vanheule}, {van Santen}, {Veenkamp},
  {Vehring}, {Voge}, {Vraeghe}, {Walck}, {Wallace}, {Wallraff}, {Wandkowsky},
  {Weaver}, {Wendt}, {Westerhoff}, {Whelan}, {Wiebe}, {Wiebusch}, {Wille},
  {Williams}, {Wills}, {Wissing}, {Wolf}, {Wood}, {Woschnagg}, {Xu}, {Xu},
  {Xu}, {Yanez}, {Yodh}, {Yoshida}, {Zoll}, \& {IceCube
  Collaboration}}]{2016ApJ...826..220A}
{Aartsen}, M.~G., {Abraham}, K., {Ackermann}, M., {et~al.} 2016, \apj, 826, 220

\bibitem[{{Abbasi} {et~al.}(2010){Abbasi}, {Abdou}, {Abu-Zayyad}, {Adams},
  {Aguilar}, {Ahlers}, {Andeen}, {Auffenberg}, {Bai}, {Baker}, \&
  et~al.}]{2010ApJ...718L.194A}
{Abbasi}, R., {Abdou}, Y., {Abu-Zayyad}, T., {et~al.} 2010, \apjl, 718, L194

\bibitem[{{Abbasi} {et~al.}(2011){Abbasi}, {Abdou}, {Abu-Zayyad}, {Adams},
  {Aguilar}, {Ahlers}, {Altmann}, {Andeen}, {Auffenberg}, {Bai}, \&
  et~al.}]{2011ApJ...740...16A}
---. 2011, \apj, 740, 16

\bibitem[{{Abbasi} {et~al.}(2012){Abbasi}, {Abdou}, {Abu-Zayyad}, {Ackermann},
  {Adams}, {Aguilar}, {Ahlers}, {Allen}, {Altmann}, {Andeen}, \&
  et~al.}]{2012ApJ...746...33A}
---. 2012, \apj, 746, 33

\bibitem[{{Abdo} {et~al.}(2008){Abdo}, {Allen}, {Aune}, {Berley}, {Blaufuss},
  {Casanova}, {Chen}, {Dingus}, {Ellsworth}, {Fleysher}, {Fleysher},
  {Gonzalez}, {Goodman}, {Hoffman}, {H{\"u}ntemeyer}, {Kolterman}, {Lansdell},
  {Linnemann}, {McEnery}, {Mincer}, {Nemethy}, {Noyes}, {Pretz}, {Ryan},
  {Parkinson}, {Shoup}, {Sinnis}, {Smith}, {Sullivan}, {Vasileiou}, {Walker},
  {Williams}, \& {Yodh}}]{2008PhRvL.101v1101A}
{Abdo}, A.~A., {Allen}, B., {Aune}, T., {et~al.} 2008, Physical Review Letters,
  101, 221101

\bibitem[{{Abdo} {et~al.}(2009){Abdo}, {Allen}, {Aune}, {Berley}, {Casanova},
  {Chen}, {Dingus}, {Ellsworth}, {Fleysher}, {Fleysher}, {Gonzalez}, {Goodman},
  {Hoffman}, {Hopper}, {H{\"u}ntemeyer}, {Kolterman}, {Lansdell}, {Linnemann},
  {McEnery}, {Mincer}, {Nemethy}, {Noyes}, {Pretz}, {Ryan}, {Parkinson},
  {Shoup}, {Sinnis}, {Smith}, {Sullivan}, {Vasileiou}, {Walker}, {Williams}, \&
  {Yodh}}]{2009ApJ...698.2121A}
{Abdo}, A.~A., {Allen}, B.~T., {Aune}, T., {et~al.} 2009, \apj, 698, 2121

\bibitem[{{Abdollahi} {et~al.}(2017){Abdollahi}, {Ackermann}, {Ajello},
  {Albert}, {Atwood}, {Baldini}, {Barbiellini}, {Bellazzini}, {Bissaldi},
  {Bloom}, {Bonino}, {Bottacini}, {Brandt}, {Bruel}, {Buson}, {Caragiulo},
  {Cavazzuti}, {Chekhtman}, {Ciprini}, {Costanza}, {Cuoco}, {Cutini},
  {D'Ammando}, {de Palma}, {Desiante}, {Digel}, {Di Lalla}, {Di Mauro}, {Di
  Venere}, {Donaggio}, {Drell}, {Favuzzi}, {Focke}, {Fukazawa}, {Funk},
  {Fusco}, {Gargano}, {Gasparrini}, {Giglietto}, {Giordano}, {Giroletti},
  {Green}, {Guiriec}, {Harding}, {Jogler}, {J{\'o}hannesson}, {Kamae}, {Kuss},
  {Larsson}, {Latronico}, {Li}, {Longo}, {Loparco}, {Lubrano}, {Magill},
  {Malyshev}, {Manfreda}, {Mazziotta}, {Meehan}, {Michelson}, {Mitthumsiri},
  {Mizuno}, {Moiseev}, {Monzani}, {Morselli}, {Negro}, {Nuss}, {Ohsugi},
  {Omodei}, {Paneque}, {Perkins}, {Pesce-Rollins}, {Piron}, {Pivato},
  {Principe}, {Rain{\`o}}, {Rando}, {Razzano}, {Reimer}, {Reimer}, {Sgr{\`o}},
  {Simone}, {Siskind}, {Spada}, {Spandre}, {Spinelli}, {Strong}, {Tajima},
  {Thayer}, {Torres}, {Troja}, {Vandenbroucke}, {Zaharijas}, {Zimmer}, \&
  {Fermi-LAT Collaboration}}]{2017PhRvL.118i1103A}
{Abdollahi}, S., {Ackermann}, M., {Ajello}, M., {et~al.} 2017, Physical Review
  Letters, 118, 091103

\bibitem[{{Abeysekara} {et~al.}(2014){Abeysekara}, {Alfaro}, {Alvarez},
  {{\'A}lvarez}, {Arceo}, {Arteaga-Vel{\'a}zquez}, {Ayala Solares}, {Barber},
  {Baughman}, {Bautista-Elivar}, {Belmont}, {BenZvi}, {Berley}, {Bonilla
  Rosales}, {Braun}, {Caballero-Mora}, {Carrami{\~n}ana}, {Castillo}, {Cotti},
  {Cotzomi}, {de la Fuente}, {De Le{\'o}n}, {DeYoung}, {Diaz Hernandez},
  {D{\'\i}az-V{\'e}lez}, {Dingus}, {DuVernois}, {Ellsworth}, {Fiorino},
  {Fraija}, {Galindo}, {Garfias}, {Gonz{\'a}lez}, {Goodman}, {Gussert},
  {Hampel-Arias}, {Harding}, {H{\"u}ntemeyer}, {Hui}, {Imran}, {Iriarte},
  {Karn}, {Kieda}, {Kunde}, {Lara}, {Lauer}, {Lee}, {Lennarz}, {Le{\'o}n
  Vargas}, {Linnemann}, {Longo}, {Luna-Garc{\'\i}a}, {Malone}, {Marinelli},
  {Marinelli}, {Martinez}, {Martinez}, {Mart{\'\i}nez-Castro}, {Matthews},
  {McEnery}, {Mendoza Torres}, {Miranda-Romagnoli}, {Moreno}, {Mostaf{\'a}},
  {Nellen}, {Newbold}, {Noriega-Papaqui}, {Oceguera-Becerra}, {Patricelli},
  {Pelayo}, {P{\'e}rez-P{\'e}rez}, {Pretz}, {Rivi{\`e}re}, {Rosa-Gonz{\'a}lez},
  {Ruiz-Velasco}, {Ryan}, {Salazar}, {Salesa Greus}, {Sandoval}, {Schneider},
  {Sinnis}, {Smith}, {Sparks Woodle}, {Springer}, {Taboada}, {Toale},
  {Tollefson}, {Torres}, {Ukwatta}, {Villase{\~n}or}, {Weisgarber},
  {Westerhoff}, {Wisher}, {Wood}, {Yodh}, {Younk}, {Zaborov}, {Zepeda}, {Zhou},
  \& {HAWC Collaboration}}]{2014ApJ...796..108A}
{Abeysekara}, A.~U., {Alfaro}, R., {Alvarez}, C., {et~al.} 2014, \apj, 796, 108

\bibitem[{{Abeysekara} {et~al.}(2017){Abeysekara}, {Albert}, {Alfaro},
  {Alvarez}, {{\'A}lvarez}, {Arceo}, {Arteaga-Vel{\'a}zquez}, {Avila Rojas},
  {Ayala Solares}, {Barber}, {Bautista-Elivar}, {Becerril}, {Belmont-Moreno},
  {BenZvi}, {Berley}, {Bernal}, {Braun}, {Brisbois}, {Caballero-Mora},
  {Capistr{\'a}n}, {Carrami{\~n}ana}, {Casanova}, {Castillo}, {Cotti},
  {Cotzomi}, {Couti{\~n}o de Le{\'o}n}, {De Le{\'o}n}, {De la Fuente},
  {Dingus}, {DuVernois}, {D{\'\i}az-V{\'e}lez}, {Ellsworth}, {Engel},
  {Enr{\'\i}quez-Rivera}, {Fiorino}, {Fraija}, {Garc{\'\i}a-Gonz{\'a}lez},
  {Garfias}, {Gerhardt}, {Gonz{\'a}lez Mu{\~n}oz}, {Gonz{\'a}lez}, {Goodman},
  {Hampel-Arias}, {Harding}, {Hern{\'a}ndez}, {Hern{\'a}ndez-Almada}, {Hinton},
  {Hona}, {Hui}, {H{\"u}ntemeyer}, {Iriarte}, {Jardin-Blicq}, {Joshi},
  {Kaufmann}, {Kieda}, {Lara}, {Lauer}, {Lee}, {Lennarz}, {Vargas},
  {Linnemann}, {Longinotti}, {Luis Raya}, {Luna-Garc{\'\i}a}, {L{\'o}pez-Coto},
  {Malone}, {Marinelli}, {Martinez}, {Martinez-Castellanos},
  {Mart{\'\i}nez-Castro}, {Mart{\'\i}nez-Huerta}, {Matthews}, {Mirand
  a-Romagnoli}, {Moreno}, {Mostaf{\'a}}, {Nellen}, {Newbold}, {Nisa},
  {Noriega-Papaqui}, {Pelayo}, {Pretz}, {P{\'e}rez-P{\'e}rez}, {Ren}, {Rho},
  {Rivi{\`e}re}, {Rosa-Gonz{\'a}lez}, {Rosenberg}, {Ruiz-Velasco}, {Salazar},
  {Salesa Greus}, {Sand oval}, {Schneider}, {Schoorlemmer}, {Sinnis}, {Smith},
  {Springer}, {Surajbali}, {Taboada}, {Tibolla}, {Tollefson}, {Torres},
  {Ukwatta}, {Vianello}, {Weisgarber}, {Westerhoff}, {Wisher}, {Wood},
  {Yapici}, {Yodh}, {Younk}, {Zepeda}, {Zhou}, {Guo}, {Hahn}, {Li}, \&
  {Zhang}}]{2017Sci...358..911A}
{Abeysekara}, A.~U., {Albert}, A., {Alfaro}, R., {et~al.} 2017, Science, 358,
  911

\bibitem[{{Ackermann} {et~al.}(2013){Ackermann}, {Ajello}, {Allafort},
  {Baldini}, {Ballet}, {Barbiellini}, {Baring}, {Bastieri}, {Bechtol},
  {Bellazzini}, {Blandford}, {Bloom}, {Bonamente}, {Borgland}, {Bottacini},
  {Brandt}, {Bregeon}, {Brigida}, {Bruel}, {Buehler}, {Busetto}, {Buson},
  {Caliandro}, {Cameron}, {Caraveo}, {Casandjian}, {Cecchi}, {{\c{C}}elik},
  {Charles}, {Chaty}, {Chaves}, {Chekhtman}, {Cheung}, {Chiang}, {Chiaro},
  {Cillis}, {Ciprini}, {Claus}, {Cohen-Tanugi}, {Cominsky}, {Conrad}, {Corbel},
  {Cutini}, {D'Ammando}, {de Angelis}, {de Palma}, {Dermer}, {do Couto e
  Silva}, {Drell}, {Drlica-Wagner}, {Falletti}, {Favuzzi}, {Ferrara},
  {Franckowiak}, {Fukazawa}, {Funk}, {Fusco}, {Gargano}, {Germani},
  {Giglietto}, {Giommi}, {Giordano}, {Giroletti}, {Glanzman}, {Godfrey},
  {Grenier}, {Grondin}, {Grove}, {Guiriec}, {Hadasch}, {Hanabata}, {Harding},
  {Hayashida}, {Hayashi}, {Hays}, {Hewitt}, {Hill}, {Hughes}, {Jackson},
  {Jogler}, {J{\'o}hannesson}, {Johnson}, {Kamae}, {Kataoka}, {Katsuta},
  {Kn{\"o}dlseder}, {Kuss}, {Lande}, {Larsson}, {Latronico}, {Lemoine-Goumard},
  {Longo}, {Loparco}, {Lovellette}, {Lubrano}, {Madejski}, {Massaro}, {Mayer},
  {Mazziotta}, {McEnery}, {Mehault}, {Michelson}, {Mignani}, {Mitthumsiri},
  {Mizuno}, {Moiseev}, {Monzani}, {Morselli}, {Moskalenko}, {Murgia},
  {Nakamori}, {Nemmen}, {Nuss}, {Ohno}, {Ohsugi}, {Omodei}, {Orienti},
  {Orlando}, {Ormes}, {Paneque}, {Perkins}, {Pesce-Rollins}, {Piron}, {Pivato},
  {Rain{\`o}}, {Rando}, {Razzano}, {Razzaque}, {Reimer}, {Reimer}, {Ritz},
  {Romoli}, {S{\'a}nchez-Conde}, {Schulz}, {Sgr{\`o}}, {Simeon}, {Siskind},
  {Smith}, {Spandre}, {Spinelli}, {Stecker}, {Strong}, {Suson}, {Tajima},
  {Takahashi}, {Takahashi}, {Tanaka}, {Thayer}, {Thayer}, {Thompson},
  {Thorsett}, {Tibaldo}, {Tibolla}, {Tinivella}, {Troja}, {Uchiyama}, {Usher},
  {Vandenbroucke}, {Vasileiou}, {Vianello}, {Vitale}, {Waite}, {Werner},
  {Winer}, {Wood}, {Wood}, {Yamazaki}, {Yang}, \&
  {Zimmer}}]{2013Sci...339..807A}
{Ackermann}, M., {Ajello}, M., {Allafort}, A., {et~al.} 2013, Science, 339, 807

\bibitem[{{Adriani} {et~al.}(2011){Adriani}, {Barbarino}, {Bazilevskaya},
  {Bellotti}, {Boezio}, {Bogomolov}, {Bonechi}, {Bongi}, {Bonvicini},
  {Borisov}, {Bottai}, {Bruno}, {Cafagna}, {Campana}, {Carbone}, {Carlson},
  {Casolino}, {Castellini}, {Consiglio}, {De Pascale}, {De Santis}, {De
  Simone}, {Di Felice}, {Galper}, {Gillard}, {Grishantseva}, {Jerse},
  {Karelin}, {Koldashov}, {Krutkov}, {Kvashnin}, {Leonov}, {Malakhov},
  {Malvezzi}, {Marcelli}, {Mayorov}, {Menn}, {Mikhailov}, {Mocchiutti},
  {Monaco}, {Mori}, {Nikonov}, {Osteria}, {Palma}, {Papini}, {Pearce},
  {Picozza}, {Pizzolotto}, {Ricci}, {Ricciarini}, {Rossetto}, {Sarkar},
  {Simon}, {Sparvoli}, {Spillantini}, {Stozhkov}, {Vacchi}, {Vannuccini},
  {Vasilyev}, {Voronov}, {Yurkin}, {Wu}, {Zampa}, {Zampa}, \&
  {Zverev}}]{2011Sci...332...69A}
{Adriani}, O., {Barbarino}, G.~C., {Bazilevskaya}, G.~A., {et~al.} 2011,
  Science, 332, 69

\bibitem[{{Aguilar} {et~al.}(2015{\natexlab{a}}){Aguilar}, {Aisa}, {Alpat},
  {Alvino}, {Ambrosi}, {Andeen}, {Arruda}, {Attig}, {Azzarello}, {Bachlechner},
  \& et~al.}]{2015PhRvL.115u1101A}
{Aguilar}, M., {Aisa}, D., {Alpat}, B., {et~al.} 2015{\natexlab{a}}, Physical
  Review Letters, 115, 211101

\bibitem[{{Aguilar} {et~al.}(2015{\natexlab{b}}){Aguilar}, {Aisa}, {Alpat},
  {Alvino}, {Ambrosi}, {Andeen}, {Arruda}, {Attig}, {Azzarello}, {Bachlechner},
  \& et~al.}]{2015PhRvL.114q1103A}
---. 2015{\natexlab{b}}, Physical Review Letters, 114, 171103

\bibitem[{{Aguilar} {et~al.}(2017){Aguilar}, {Ali Cavasonza}, {Alpat},
  {Ambrosi}, {Arruda}, {Attig}, {Aupetit}, {Azzarello}, {Bachlechner}, {Barao},
  \& et~al.}]{2017PhRvL.119y1101A}
{Aguilar}, M., {Ali Cavasonza}, L., {Alpat}, B., {et~al.} 2017, Physical Review
  Letters, 119, 251101

\bibitem[{{Aguilar} {et~al.}(2019){Aguilar}, {Ali Cavasonza}, {Ambrosi},
  {Arruda}, {Attig}, {Azzarello}, {Bachlechner}, {Barao}, {Barrau}, {Barrin},
  {Bartoloni}, {Basara}, {Ba{\textcommabelow s}e{\v{g}}mez-du Pree},
  {Battiston}, {Becker}, {Behlmann}, {Beischer}, {Berdugo}, {Bertucci},
  {Bindi}, {de Boer}, {Bollweg}, {Borgia}, {Boschini}, {Bourquin}, {Bueno},
  {Burger}, {Burger}, {Cai}, {Capell}, {Caroff}, {Casaus}, {Castellini},
  {Cervelli}, {Chang}, {Chen}, {Chen}, {Chen}, {Cheng}, {Chou}, {Choutko},
  {Chung}, {Clark}, {Coignet}, {Consolandi}, {Contin}, {Corti}, {Crispoltoni},
  {Cui}, {Dadzie}, {Dai}, {Datta}, {Delgado}, {Della Torre}, {Demirk{\"o}z},
  {Derome}, {Di Falco}, {Dimiccoli}, {D{\'\i}az}, {von Doetinchem}, {Dong},
  {Donnini}, {Duranti}, {Egorov}, {Eline}, {Eronen}, {Feng}, {Fiandrini},
  {Fisher}, {Formato}, {Galaktionov}, {Garc{\'\i}a-L{\'o}pez}, {Gargiulo},
  {Gast}, {Gebauer}, {Gervasi}, {Giovacchini}, {G{\'o}mez-Coral}, {Gong},
  {Goy}, {Grabski}, {Grandi}, {Graziani}, {Guo}, {Haino}, {Han}, {He}, {Heil},
  {Hsieh}, {Huang}, {Huang}, {Incagli}, {Jia}, {Jinchi}, {Kanishev}, {Khiali},
  {Kirn}, {Konak}, {Kounina}, {Kounine}, {Koutsenko}, {Kulemzin}, {La Vacca},
  {Laudi}, {Laurenti}, {Lazzizzera}, {Lebedev}, {Lee}, {Lee}, {Leluc}, {Li},
  {Li}, {Li}, {Li}, {Light}, {Lin}, {Lippert}, {Liu}, {Liu}, {Liu}, {Lu}, {Lu},
  {Luebelsmeyer}, {Luo}, {Luo}, {Luo}, {Lyu}, {Machate}, {Ma{\~n}{\'a}},
  {Mar{\'\i}n}, {Martin}, {Mart{\'\i}nez}, {Masi}, {Maurin}, {Menchaca-Rocha},
  {Meng}, {Mo}, {Molero}, {Mott}, {Mussolin}, {Nelson}, {Ni}, {Nikonov},
  {Nozzoli}, {Oliva}, {Orcinha}, {Palermo}, {Palmonari}, {Paniccia}, {Pashnin},
  {Pauluzzi}, {Pensotti}, {Perrina}, {Phan}, {Picot-Clemente}, {Plyaskin},
  {Pohl}, {Poireau}, {Popkow}, {Quadrani}, {Qi}, {Qin}, {Qu}, {Rancoita},
  {Rapin}, {Conde}, {Rosier-Lees}, {Rozhkov}, {Rozza}, {Sagdeev}, {Solano},
  {Schael}, {Schmidt}, {Schulz von Dratzig}, {Schwering}, {Seo}, {Shan}, {Shi},
  {Siedenburg}, {Song}, {Sun}, {Tacconi}, {Tang}, {Tang}, {Tian}, {Ting},
  {Ting}, {Tomassetti}, {Torsti}, {Urban}, {Vagelli}, {Valente}, {Valtonen},
  {V{\'a}zquez Acosta}, {Vecchi}, {Velasco}, {Vialle}, {Viz{\'a}n}, {Wang},
  {Wang}, {Wang}, {Wang}, {Wang}, {Wang}, {Wei}, {Weng}, {Wu}, {Xiong}, {Xu},
  {Yan}, {Yang}, {Yi}, {Yu}, {Yu}, {Zannoni}, {Zeissler}, {Zhang}, {Zhang},
  {Zhang}, {Zhang}, {Zhao}, {Zheng}, {Zhuang}, {Zhukov}, {Zichichi},
  {Zimmermann}, {Zuccon}, \& {AMS Collaboration}}]{2019PhRvL.122d1102A}
{Aguilar}, M., {Ali Cavasonza}, L., {Ambrosi}, G., {et~al.} 2019, \prl, 122,
  041102

\bibitem[{{Aharonian} {et~al.}(2008){Aharonian}, {Akhperjanian}, {Barres de
  Almeida}, {Bazer-Bachi}, {Becherini}, {Behera}, {Benbow}, {Bernl{\"o}hr},
  {Boisson}, {Bochow}, {Borrel}, {Braun}, {Brion}, {Brucker}, {Brun},
  {B{\"u}hler}, {Bulik}, {B{\"u}sching}, {Boutelier}, {Carrigan}, {Chadwick},
  {Charbonnier}, {Chaves}, {Cheesebrough}, {Chounet}, {Clapson}, {Coignet},
  {Costamante}, {Dalton}, {Degrange}, {Deil}, {Dickinson}, {Djannati-Ata{\"i}},
  {Domainko}, {Drury}, {Dubois}, {Dubus}, {Dyks}, {Dyrda}, {Egberts},
  {Emmanoulopoulos}, {Espigat}, {Farnier}, {Feinstein}, {Fiasson},
  {F{\"o}rster}, {Fontaine}, {F{\"u}{\ss}ling}, {Gabici}, {Gallant},
  {G{\'e}rard}, {Giebels}, {Glicenstein}, {Gl{\"u}ck}, {Goret}, {Vivier},
  {V{\"o}lk}, {Volpe}, {Wagner}, {Ward}, {Zdziarski}, \&
  {Zech}}]{2008PhRvL.101z1104A}
{Aharonian}, F., {Akhperjanian}, A.~G., {Barres de Almeida}, U., {et~al.} 2008,
  Physical Review Letters, 101, 261104

\bibitem[{{Ahlers}(2014)}]{2014PhRvL.112b1101A}
{Ahlers}, M. 2014, \prl, 112, 021101

\bibitem[{{Ahn} {et~al.}(2010){Ahn}, {Allison}, {Bagliesi}, {Beatty},
  {Bigongiari}, {Childers}, {Conklin}, {Coutu}, {DuVernois}, {Ganel}, {Han},
  {Jeon}, {Kim}, {Lee}, {Lutz}, {Maestro}, {Malinin}, {Marrocchesi}, {Minnick},
  {Mognet}, {Nam}, {Nam}, {Nutter}, {Park}, {Park}, {Seo}, {Sina}, {Wu},
  {Yang}, {Yoon}, {Zei}, \& {Zinn}}]{2010ApJ...714L..89A}
{Ahn}, H.~S., {Allison}, P., {Bagliesi}, M.~G., {et~al.} 2010, \apjl, 714, L89

\bibitem[{{Alemanno} {et~al.}(2021){Alemanno}, {An}, {Azzarello}, {Barbato},
  {Bernardini}, {Bi}, {Cai}, {Catanzani}, {Chang}, {Chen}, {Chen}, {Chen},
  {Cui}, {Cui}, {Cui}, {Dai}, {D'Amone}, {de Benedittis}, {de Mitri}, {de
  Palma}, {Deliyergiyev}, {di Santo}, {Dong}, {Dong}, {Donvito}, {Droz},
  {Duan}, {Duan}, {D'Urso}, {Fan}, {Fan}, {Fang}, {Fang}, {Feng}, {Feng},
  {Fusco}, {Gao}, {Gargano}, {Gong}, {Gong}, {Guo}, {Guo}, {Guo}, {Han}, {Hu},
  {Huang}, {Huang}, {Huang}, {Ionica}, {Jiang}, {Kong}, {Kotenko}, {Kyratzis},
  {Lei}, {Li}, {Li}, {Li}, {Li}, {Liang}, {Liu}, {Liu}, {Liu}, {Liu}, {Liu},
  {Liu}, {Loparco}, {Luo}, {Ma}, {Ma}, {Ma}, {Ma}, {Marsella}, {Mazziotta},
  {Mo}, {Niu}, {Pan}, {Parenti}, {Peng}, {Peng}, {Perrina}, {Qiao}, {Rao},
  {Ruina}, {Salinas}, {Shang}, {Shen}, {Shen}, {Shen}, {Silveri}, {Song},
  {Stolpovskiy}, {Su}, {Su}, {Sun}, {Surdo}, {Teng}, {Tykhonov}, {Wang},
  {Wang}, {Wang}, {Wang}, {Wang}, {Wang}, {Wang}, {Wang}, {Wang}, {Wei}, {Wei},
  {Wei}, {Wen}, {Wu}, {Wu}, {Wu}, {Wu}, {Wu}, {Xia}, {Xu}, {Xu}, {Xu}, {Xu},
  {Xue}, {Yang}, {Yang}, {Yang}, {Yao}, {Yu}, {Yuan}, {Yuan}, {Yue}, {Zang},
  {Zhang}, {Zhang}, {Zhang}, {Zhang}, {Zhang}, {Zhang}, {Zhang}, {Zhang},
  {Zhang}, {Zhang}, {Zhao}, {Zhao}, {Zhao}, {Zhou}, {Zhu}, \& {Dampe
  Collaboration}}]{2021PhRvL.126t1102A}
{Alemanno}, F., {An}, Q., {Azzarello}, P., {et~al.} 2021, \prl, 126, 201102

\bibitem[{{Alvarez} {et~al.}(2001){Alvarez}, {Aparici}, {May}, \&
  {Reich}}]{2001A&A...372..636A}
{Alvarez}, H., {Aparici}, J., {May}, J., \& {Reich}, P. 2001, \aap, 372, 636

\bibitem[{{Amenomori} {et~al.}(2005){Amenomori}, {Ayabe}, {Cui},
  {Danzengluobu}, {Ding}, {Ding}, {Feng}, {Feng}, {Gao}, {Geng}, {Guo}, {He},
  {He}, {Hibino}, {Hotta}, {Hu}, {Hu}, {Huang}, {Huang}, {Jia}, {Kajino},
  {Kasahara}, {Katayose}, {Kato}, {Kawata}, {Labaciren}, {Le}, {Li}, {Lu},
  {Lu}, {Meng}, {Mizutani}, {Mu}, {Munakata}, {Nagai}, {Nanjo}, {Nishizawa},
  {Ohnishi}, {Ohta}, {Onuma}, {Ouchi}, {Ozawa}, {Ren}, {Saito}, {Sakata},
  {Sasaki}, {Shibata}, {Shiomi}, {Shirai}, {Sugimoto}, {Takita}, {Tan},
  {Tateyama}, {Torii}, {Tsuchiya}, {Udo}, {Utsugi}, {Wang}, {Wang}, {Wang},
  {Wang}, {Wu}, {Xue}, {Yamamoto}, {Yan}, {Yang}, {Yasue}, {Ye}, {Yu}, {Yuan},
  {Yuda}, {Zhang}, {Zhang}, {Zhang}, {Zhang}, {Zhang}, {Zhang}, {Zhaxisangzhu},
  {Zhou}, \& {Tibet As{$\gamma$} Collaboration}}]{2005ApJ...626L..29A}
{Amenomori}, M., {Ayabe}, S., {Cui}, S.~W., {et~al.} 2005, \apjl, 626, L29

\bibitem[{{Amenomori} {et~al.}(2006){Amenomori}, {Ayabe}, {Bi}, {Chen}, {Cui},
  {Danzengluobu}, {Ding}, {Ding}, {Feng}, {Feng}, {Feng}, {Gao}, {Geng}, {Guo},
  {He}, {He}, {Hibino}, {Hotta}, {Hu}, {Hu}, {Huang}, {Huang}, {Jia}, {Kajino},
  {Kasahara}, {Katayose}, {Kato}, {Kawata}, {Labaciren}, {Le}, {Li}, {Li},
  {Lou}, {Lu}, {Lu}, {Meng}, {Mizutani}, {Mu}, {Munakata}, {Nagai}, {Nanjo},
  {Nishizawa}, {Ohnishi}, {Ohta}, {Onuma}, {Ouchi}, {Ozawa}, {Ren}, {Saito},
  {Saito}, {Sakata}, {Sako}, {Sasaki}, {Shibata}, {Shiomi}, {Shirai},
  {Sugimoto}, {Takita}, {Tan}, {Tateyama}, {Torii}, {Tsuchiya}, {Udo}, {Wang},
  {Wang}, {Wang}, {Wang}, {Wu}, {Xue}, {Yamamoto}, {Yan}, {Yang}, {Yasue},
  {Ye}, {Yu}, {Yuan}, {Yuda}, {Zhang}, {Zhang}, {Zhang}, {Zhang}, {Zhang},
  {Zhang}, {Zhaxisangzhu}, {Zhou}, \& {Tibet AS{\ensuremath{\gamma}}
  Collaboration}}]{2006Sci...314..439A}
{Amenomori}, M., {Ayabe}, S., {Bi}, X.~J., {et~al.} 2006, Science, 314, 439

\bibitem[{{Amenomori} {et~al.}(2010){Amenomori}, {Bi}, {Chen}, {Cui},
  {Danzengluobu}, {Ding}, {Ding}, {Fan}, {Feng}, {Feng}, {Feng}, {Gao}, {Geng},
  {Gou}, {Guo}, {He}, {He}, {Hibino}, {Hotta}, {Hu}, {Hu}, {Huang}, {Huang},
  {Jia}, {Jiang}, {Kajino}, {Kasahara}, {Katayose}, {Kato}, {Kawata},
  {Labaciren}, {Le}, {Li}, {Li}, {Li}, {Liu}, {Lou}, {Lu}, {Meng}, {Mizutani},
  {Mu}, {Munakata}, {Nagai}, {Nanjo}, {Nishizawa}, {Ohnishi}, {Ohta}, {Ozawa},
  {Saito}, {Saito}, {Sakata}, {Sako}, {Shibata}, {Shiomi}, {Shirai},
  {Sugimoto}, {Takita}, {Tan}, {Tateyama}, {Torii}, {Tsuchiya}, {Udo}, {Wang},
  {Wang}, {Wang}, {Wang}, {Wu}, {Xue}, {Yamamoto}, {Yan}, {Yang}, {Yasue},
  {Ye}, {Yu}, {Yuan}, {Yuda}, {Zhang}, {Zhang}, {Zhang}, {Zhang}, {Zhang},
  {Zhang}, {Zhang}, {Zhaxisangzhu}, {Zhou}, \& {Tibet AS{\ensuremath{\gamma}}
  Collaboration}}]{2010ApJ...711..119A}
{Amenomori}, M., {Bi}, X.~J., {Chen}, D., {et~al.} 2010, \apj, 711, 119

\bibitem[{{Amenomori} {et~al.}(2017){Amenomori}, {Bi}, {Chen}, {Chen}, {Chen},
  {Cui}, {Danzengluobu}, {Ding}, {Feng}, {Feng}, {Feng}, {Gou}, {Guo}, {He},
  {He}, {Hibino}, {Hotta}, {Hu}, {Hu}, {Huang}, {Jia}, {Jiang}, {Kajino},
  {Kasahara}, {Katayose}, {Kato}, {Kawata}, {Kozai}, {Labaciren}, {Le}, {Li},
  {Li}, {Li}, {Liu}, {Liu}, {Liu}, {Lu}, {Meng}, {Miyazaki}, {Mizutani},
  {Munakata}, {Nakajima}, {Nakamura}, {Nanjo}, {Nishizawa}, {Niwa}, {Ohnishi},
  {Ohta}, {Ozawa}, {Qian}, {Qu}, {Saito}, {Saito}, {Sakata}, {Sako}, {Shao},
  {Shibata}, {Shiomi}, {Shirai}, {Sugimoto}, {Takita}, {Tan}, {Tateyama},
  {Torii}, {Tsuchiya}, {Udo}, {Wang}, {Wu}, {Xue}, {Yamamoto}, {Yamauchi},
  {Yang}, {Yuan}, {Yuda}, {Zhai}, {Zhang}, {Zhang}, {Zhang}, {Zhang}, {Zhang},
  {Zhang}, {Zhaxisangzhu}, {Zhou}, \& {Tibet AS{\ensuremath{\gamma}}
  Collaboration}}]{2017ApJ...836..153A}
---. 2017, \apj, 836, 153

\bibitem[{{An} {et~al.}(2019){An}, {Asfandiyarov}, {Azzarello}, {Bernardini},
  {Bi}, {Cai}, {Chang}, {Chen}, {Chen}, {Chen}, {Chen}, {Cui}, {Cui}, {Dai},
  {D'Amone}, {De Benedittis}, {De Mitri}, {Di Santo}, {Ding}, {Dong}, {Dong},
  {Dong}, {Donvito}, {Droz}, {Duan}, {Duan}, {D'Urso}, {Fan}, {Fan}, {Fang},
  {Feng}, {Feng}, {Fusco}, {Gallo}, {Gan}, {Gao}, {Gargano}, {Gong}, {Gong},
  {Guo}, {Guo}, {Guo}, {Han}, {Hu}, {Huang}, {Huang}, {Huang}, {Ionica},
  {Jiang}, {Jin}, {Kong}, {Lei}, {Li}, {Li}, {Li}, {Li}, {Li}, {Liang},
  {Liang}, {Liao}, {Liu}, {Liu}, {Liu}, {Liu}, {Liu}, {Liu}, {Loparco}, {Luo},
  {Ma}, {Ma}, {Ma}, {Ma}, {Ma}, {Marsella}, {Mazziotta}, {Mo}, {Niu}, {Pan},
  {Peng}, {Peng}, {Qiao}, {Rao}, {Salinas}, {Shang}, {Shen}, {Shen}, {Shen},
  {Song}, {Su}, {Su}, {Sun}, {Surdo}, {Teng}, {Tykhonov}, {Vitillo}, {Wang},
  {Wang}, {Wang}, {Wang}, {Wang}, {Wang}, {Wang}, {Wang}, {Wang}, {Wang},
  {Wang}, {Wang}, {Wang}, {Wei}, {Wei}, {Wei}, {Wen}, {Wu}, {Wu}, {Wu}, {Wu},
  {Wu}, {Xi}, {Xia}, {Xu}, {Xu}, {Xu}, {Xu}, {Xue}, {Yang}, {Yang}, {Yang},
  {Yang}, {Yao}, {Yu}, {Yuan}, {Yue}, {Zang}, {Zhang}, {Zhang}, {Zhang},
  {Zhang}, {Zhang}, {Zhang}, {Zhang}, {Zhang}, {Zhang}, {Zhang}, {Zhang},
  {Zhang}, {Zhang}, {Zhao}, {Zhao}, {Zhao}, {Zhou}, {Zhou}, {Zhu}, {Zhu}, \&
  {Zimmer}}]{2019SciA....5.3793A}
{An}, Q., {Asfandiyarov}, R., {Azzarello}, P., {et~al.} 2019, Science Advances,
  5, eaax3793

\bibitem[{{Archer} {et~al.}(2018){Archer}, {Benbow}, {Bird}, {Brose},
  {Buchovecky}, {Buckley}, {Bugaev}, {Connolly}, {Cui}, {Daniel}, {Feng},
  {Finley}, {Fortson}, {Furniss}, {Gillanders}, {H{\"u}tten}, {Hanna},
  {Hervet}, {Holder}, {Hughes}, {Humensky}, {Johnson}, {Kaaret}, {Kar},
  {Kelley-Hoskins}, {Kertzman}, {Kieda}, {Krause}, {Krennrich}, {Kumar},
  {Lang}, {Lin}, {Maier}, {McArthur}, {Moriarty}, {Mukherjee}, {O'Brien},
  {Ong}, {Otte}, {Petrashyk}, {Pohl}, {Pueschel}, {Quinn}, {Ragan}, {Reynolds},
  {Richards}, {Roache}, {Rulten}, {Sadeh}, {Santander}, {Sembroski}, {Staszak},
  {Sushch}, {Wakely}, {Wells}, {Wilcox}, {Wilhelm}, {Williams}, {Williamson},
  {Zitzer}, \& {VERITAS Collaboration}}]{2018PhRvD..98f2004A}
{Archer}, A., {Benbow}, W., {Bird}, R., {et~al.} 2018, \prd, 98, 062004

\bibitem[{{Aschenbach}(1998)}]{1998Natur.396..141A}
{Aschenbach}, B. 1998, \nat, 396, 141

\bibitem[{{Atkin} {et~al.}(2017){Atkin}, {Bulatov}, {Dorokhov}, {Gorbunov},
  {Filippov}, {Grebenyuk}, {Karmanov}, {Kovalev}, {Kudryashov}, {Kurganov},
  {Merkin}, {Panov}, {Podorozhny}, {Polkov}, {Porokhovoy}, {Shumikhin},
  {Sveshnikova}, {Tkachenko}, {Tkachev}, {Turundaevskiy}, {Vasiliev}, \&
  {Voronin}}]{2017JCAP...07..020A}
{Atkin}, E., {Bulatov}, V., {Dorokhov}, V., {et~al.} 2017, \jcap, 7, 020

\bibitem[{{Atkin} {et~al.}(2018){Atkin}, {Bulatov}, {Dorokhov}, {Gorbunov},
  {Filippov}, {Grebenyuk}, {Karmanov}, {Kovalev}, {Kudryashov}, {Kurganov},
  {Merkin}, {Panov}, {Podorozhny}, {Polkov}, {Porokhovoy}, {Shumikhin},
  {Tkachenko}, {Tkachev}, {Turundaevskiy}, {Vasiliev}, \&
  {Voronin}}]{2018JETPL.108....5A}
---. 2018, Soviet Journal of Experimental and Theoretical Physics Letters, 108,
  5

\bibitem[{{Atoyan} {et~al.}(1995){Atoyan}, {Aharonian}, \&
  {V{\"o}lk}}]{1995PhRvD..52.3265A}
{Atoyan}, A.~M., {Aharonian}, F.~A., \& {V{\"o}lk}, H.~J. 1995, \prd, 52, 3265

\bibitem[{{Bartoli} {et~al.}(2013){Bartoli}, {Bernardini}, {Bi}, {Bolognino},
  {Branchini}, {Budano}, {Calabrese Melcarne}, {Camarri}, {Cao}, {Cardarelli},
  {Catalanotti}, {Chen}, {Chen}, {Creti}, {Cui}, {Dai}, {D'Amone},
  {Danzengluobu}, {De Mitri}, {D'Ettorre Piazzoli}, {Di Girolamo}, {Di
  Sciascio}, {Feng}, {Feng}, {Feng}, {Gou}, {Guo}, {He}, {Hu}, {Hu},
  {Iacovacci}, {Iuppa}, {Jia}, {Labaciren}, {Li}, {Liguori}, {Liu}, {Liu},
  {Liu}, {Lu}, {Ma}, {Mancarella}, {Mari}, {Marsella}, {Martello},
  {Mastroianni}, {Montini}, {Ning}, {Panareo}, {Panico}, {Perrone}, {Pistilli},
  {Ruggieri}, {Salvini}, {Santonico}, {Sbano}, {Shen}, {Sheng}, {Shi}, {Surdo},
  {Tan}, {Vallania}, {Vernetto}, {Vigorito}, {Wang}, {Wu}, {Wu}, {Xue}, {Yan},
  {Yang}, {Yang}, {Yao}, {Yuan}, {Zha}, {Zhang}, {Zhang}, {Zhang}, {Zhang},
  {Zhaxiciren}, {Zhaxisangzhu}, {Zhou}, {Zhu}, {Zhu}, \&
  {Zizzi}}]{2013PhRvD..88h2001B}
{Bartoli}, B., {Bernardini}, P., {Bi}, X.~J., {et~al.} 2013, \prd, 88, 082001

\bibitem[{{Bartoli} {et~al.}(2015){Bartoli}, {Bernardini}, {Bi}, {Cao},
  {Catalanotti}, {Chen}, {Chen}, {Cui}, {Dai}, {D'Amone}, {Danzengluobu}, {De
  Mitri}, {D'Ettorre Piazzoli}, {Di Girolamo}, {Di Sciascio}, {Feng}, {Feng},
  {Feng}, {Gao}, {Gou}, {Guo}, {He}, {Hu}, {Hu}, {Iacovacci}, {Iuppa}, {Jia},
  {Labaciren}, {Li}, {Liu}, {Liu}, {Liu}, {Lu}, {Ma}, {Ma}, {Mancarella},
  {Mari}, {Marsella}, {Mastroianni}, {Montini}, {Ning}, {Perrone}, {Pistilli},
  {Salvini}, {Santonico}, {Shen}, {Sheng}, {Shi}, {Surdo}, {Tan}, {Vallania},
  {Vernetto}, {Vigorito}, {Wang}, {Wu}, {Wu}, {Xue}, {Yang}, {Yang}, {Yao},
  {Yuan}, {Zha}, {Zhang}, {Zhang}, {Zhang}, {Zhang}, {Zhao}, {Zhaxiciren},
  {Zhaxisangzhu}, {Zhou}, {Zhu}, {Zhu}, \& {ARGO-YBJ
  Collaboration}}]{2015ApJ...809...90B}
---. 2015, \apj, 809, 90

\bibitem[{{Bernard} {et~al.}(2012){Bernard}, {Delahaye}, {Salati}, \&
  {Taillet}}]{2012A&A...544A..92B}
{Bernard}, G., {Delahaye}, T., {Salati}, P., \& {Taillet}, R. 2012, \aap, 544,
  A92

\bibitem[{{Blair} {et~al.}(2005){Blair}, {Sankrit}, \&
  {Raymond}}]{2005AJ....129.2268B}
{Blair}, W.~P., {Sankrit}, R., \& {Raymond}, J.~C. 2005, \aj, 129, 2268

\bibitem[{{Blasi}(2013)}]{2013A&ARv..21...70B}
{Blasi}, P. 2013, \aapr, 21, 70

\bibitem[{{Boschini} {et~al.}(2017){Boschini}, {Della Torre}, {Gervasi},
  {Grandi}, {J{\'o}hannesson}, {Kachelriess}, {La Vacca}, {Masi}, {Moskalenko},
  {Orlando}, {Ostapchenko}, {Pensotti}, {Porter}, {Quadrani}, {Rancoita},
  {Rozza}, \& {Tacconi}}]{2017ApJ...840..115B}
{Boschini}, M.~J., {Della Torre}, S., {Gervasi}, M., {et~al.} 2017, \apj, 840,
  115

\bibitem[{{Boulares}(1989)}]{1989ApJ...342..807B}
{Boulares}, A. 1989, \apj, 342, 807

\bibitem[{{Brisken} {et~al.}(2003){Brisken}, {Thorsett}, {Golden}, \&
  {Goss}}]{2003ApJ...593L..89B}
{Brisken}, W.~F., {Thorsett}, S.~E., {Golden}, A., \& {Goss}, W.~M. 2003,
  \apjl, 593, L89

\bibitem[{{Caraveo} {et~al.}(2001){Caraveo}, {De Luca}, {Mignani}, \&
  {Bignami}}]{2001ApJ...561..930C}
{Caraveo}, P.~A., {De Luca}, A., {Mignani}, R.~P., \& {Bignami}, G.~F. 2001,
  \apj, 561, 930

\bibitem[{{Cerri} {et~al.}(2017){Cerri}, {Gaggero}, {Vittino}, {Evoli}, \&
  {Grasso}}]{2017JCAP...10..019C}
{Cerri}, S.~S., {Gaggero}, D., {Vittino}, A., {Evoli}, C., \& {Grasso}, D.
  2017, \jcap, 2017, 019

\bibitem[{{Cha} {et~al.}(1999){Cha}, {Sembach}, \&
  {Danks}}]{1999ApJ...515L..25C}
{Cha}, A.~N., {Sembach}, K.~R., \& {Danks}, A.~C. 1999, \apjl, 515, L25

\bibitem[{{DAMPE Collaboration} {et~al.}(2017){DAMPE Collaboration}, {Ambrosi},
  {An}, {Asfandiyarov}, {Azzarello}, {Bernardini}, {Bertucci}, {Cai}, {Chang},
  {Chen}, {Chen}, {Chen}, {Chen}, {Cui}, {Cui}, {D'Amone}, {de Benedittis}, {De
  Mitri}, {di Santo}, {Dong}, {Dong}, {Dong}, {Dong}, {Donvito}, {Droz},
  {Duan}, {Duan}, {Duranti}, {D'Urso}, {Fan}, {Fan}, {Fang}, {Feng}, {Feng},
  {Fusco}, {Gallo}, {Gan}, {Gao}, {Gao}, {Gargano}, {Garrappa}, {Gong}, {Gong},
  {Guo}, {Guo}, {Hu}, {Huang}, {Huang}, {Ionica}, {Jiang}, {Jiang}, {Jin},
  {Kong}, {Lei}, {Li}, {Li}, {Li}, {Li}, {Liang}, {Liang}, {Liao}, {Liu},
  {Liu}, {Liu}, {Liu}, {Liu}, {Loparco}, {Ma}, {Ma}, {Ma}, {Ma}, {Ma}, {Ma},
  {Marsella}, {Mazziotta}, {Mo}, {Niu}, {Peng}, {Peng}, {Qiao}, {Rao},
  {Salinas}, {Shang}, {H.~Shen}, {Shen}, {Shen}, {Song}, {Su}, {Su}, {Sun},
  {Surdo}, {Teng}, {Tian}, {Tykhonov}, {Vagelli}, {Vitillo}, {Wang}, {Wang},
  {Wang}, {Wang}, {Wang}, {Wang}, {Wang}, {Wang}, {Wang}, {Wang}, {Wang},
  {Wang}, {Wen}, {Wang}, {Wei}, {Wei}, {Wei}, {Wu}, {Wu}, {Wu}, {Wu}, {Wu},
  {Xi}, {Xia}, {Xin}, {Xu}, {Xu}, {Xu}, {Xue}, {Yang}, {Yang}, {Yang}, {Yang},
  {Yao}, {Yu}, {Yuan}, {Yue}, {Zang}, {Zhang}, {Zhang}, {Zhang}, {Zhang},
  {Zhang}, {Zhang}, {Zhang}, {Zhang}, {Zhang}, {Zhang}, {Zhang}, {Zhang},
  {Zhang}, {Zhang}, {Zhang}, {Zhang}, {Zhang}, {Zhao}, {Zhao}, {Zhao}, {Zhou},
  {Zhou}, {Zhu}, {Zhu}, \& {Zimmer}}]{2017Natur.552...63D}
{DAMPE Collaboration}, {Ambrosi}, G., {An}, Q., {et~al.} 2017, \nat, 552, 63

\bibitem[{{Di Mauro} {et~al.}(2014){Di Mauro}, {Donato}, {Fornengo}, {Lineros},
  \& {Vittino}}]{2014JCAP...04..006D}
{Di Mauro}, M., {Donato}, F., {Fornengo}, N., {Lineros}, R., \& {Vittino}, A.
  2014, \jcap, 4, 6

\bibitem[{{Fang} {et~al.}(2018{\natexlab{a}}){Fang}, {Bi}, \&
  {Yin}}]{2018ApJ...854...57F}
{Fang}, K., {Bi}, X.-J., \& {Yin}, P.-F. 2018{\natexlab{a}}, \apj, 854, 57

\bibitem[{{Fang} {et~al.}(2018{\natexlab{b}}){Fang}, {Bi}, {Yin}, \&
  {Yuan}}]{2018ApJ...863...30F}
{Fang}, K., {Bi}, X.-J., {Yin}, P.-F., \& {Yuan}, Q. 2018{\natexlab{b}}, \apj,
  863, 30

\bibitem[{{Fang} {et~al.}(2017){Fang}, {Wang}, {Bi}, {Lin}, \&
  {Yin}}]{2017ApJ...836..172F}
{Fang}, K., {Wang}, B.-B., {Bi}, X.-J., {Lin}, S.-J., \& {Yin}, P.-F. 2017,
  \apj, 836, 172

\bibitem[{{Feng} {et~al.}(2016){Feng}, {Tomassetti}, \&
  {Oliva}}]{2016PhRvD..94l3007F}
{Feng}, J., {Tomassetti}, N., \& {Oliva}, A. 2016, \prd, 94, 123007

\bibitem[{{Gleeson} \& {Axford}(1968)}]{1968ApJ...154.1011G}
{Gleeson}, L.~J., \& {Axford}, W.~I. 1968, \apj, 154, 1011

\bibitem[{{Gorham} {et~al.}(1996){Gorham}, {Ray}, {Anderson}, {Kulkarni}, \&
  {Prince}}]{1996ApJ...458..257G}
{Gorham}, P.~W., {Ray}, P.~S., {Anderson}, S.~B., {Kulkarni}, S.~R., \&
  {Prince}, T.~A. 1996, \apj, 458, 257

\bibitem[{{Green}(2014)}]{2014BASI...42...47G}
{Green}, D.~A. 2014, Bulletin of the Astronomical Society of India, 42, 47

\bibitem[{{Green}(2015)}]{2015MNRAS.454.1517G}
---. 2015, \mnras, 454, 1517

\bibitem[{{Guillian} {et~al.}(2007){Guillian}, {Hosaka}, {Ishihara}, {Kameda},
  {Koshio}, {Minamino}, {Mitsuda}, {Miura}, {Moriyama}, {Nakahata}, {Namba},
  {Obayashi}, {Ogawa}, {Shiozawa}, {Suzuki}, {Takeda}, {Takeuchi}, {Yamada},
  {Higuchi}, {Ishitsuka}, {Kajita}, {Kaneyuki}, {Mitsuka}, {Nakayama},
  {Nishino}, {Okada}, {Okumura}, {Saji}, {Takenaga}, {Desai}, {Kearns},
  {Stone}, {Sulak}, {Wang}, {Goldhaber}, {Casper}, {Gajewski}, {Griskevich},
  {Kropp}, {Liu}, {Mine}, {Smy}, {Sobel}, {Vagins}, {Ganezer}, {Hill}, {Keig},
  {Scholberg}, {Walter}, {Ellsworth}, {Tasaka}, {Kibayashi}, {Learned},
  {Matsuno}, {Messier}, {Hayato}, {Ichikawa}, {Ishida}, {Ishii}, {Iwashita},
  {Kobayashi}, {Nakadaira}, {Nakamura}, {Nitta}, {Oyama}, {Totsuka}, {Suzuki},
  {Hasegawa}, {Kato}, {Maesaka}, {Nakaya}, {Nishikawa}, {Sato}, {Yamamoto},
  {Yokoyama}, {Haines}, {Dazeley}, {Hatakeyama}, {Svoboda}, {Blaufuss},
  {Goodman}, {Sullivan}, {Turcan}, {Habig}, {Fukuda}, {Itow}, {Sakuda},
  {Yoshida}, {Kim}, {Yoo}, {Okazawa}, {Ishizuka}, {Jung}, {Kato}, {Kobayashi},
  {Malek}, {Mauger}, {McGrew}, {Sharkey}, {Yanagisawa}, {Gando}, {Hasegawa},
  {Inoue}, {Shirai}, {Suzuki}, {Nishijima}, {Ishino}, {Watanabe}, {Koshiba},
  {Kielczewska}, {Berns}, {Gran}, {Shiraishi}, {Stachyra}, {Washburn},
  {Wilkes}, \& {Munakata}}]{2007PhRvD..75f2003G}
{Guillian}, G., {Hosaka}, J., {Ishihara}, K., {et~al.} 2007, \prd, 75, 062003

\bibitem[{{Guo} {et~al.}(2016){Guo}, {Tian}, \& {Jin}}]{2016ApJ...819...54G}
{Guo}, Y.-Q., {Tian}, Z., \& {Jin}, C. 2016, \apj, 819, 54

\bibitem[{{Guo} \& {Yuan}(2018)}]{2018PhRvD..97f3008G}
{Guo}, Y.-Q., \& {Yuan}, Q. 2018, \prd, 97, 063008

\bibitem[{{Iyudin} {et~al.}(1998){Iyudin}, {Sch{\"o}nfelder}, {Bennett},
  {Bloemen}, {Diehl}, {Hermsen}, {Lichti}, {van der Meulen}, {Ryan}, \&
  {Winkler}}]{1998Natur.396..142I}
{Iyudin}, A.~F., {Sch{\"o}nfelder}, V., {Bennett}, K., {et~al.} 1998, \nat,
  396, 142

\bibitem[{{Jin} {et~al.}(2016){Jin}, {Guo}, \& {Hu}}]{2016ChPhC..40a5101J}
{Jin}, C., {Guo}, Y.-Q., \& {Hu}, H.-B. 2016, Chinese Physics C, 40, 015101

\bibitem[{{Joncas} {et~al.}(1989){Joncas}, {Roger}, \&
  {Dewdney}}]{1989A&A...219..303J}
{Joncas}, G., {Roger}, R.~S., \& {Dewdney}, P.~E. 1989, \aap, 219, 303

\bibitem[{{Katsuda} {et~al.}(2009){Katsuda}, {Petre}, {Hwang}, {Yamaguchi},
  {Mori}, \& {Tsunemi}}]{2009PASJ...61S.155K}
{Katsuda}, S., {Petre}, R., {Hwang}, U., {et~al.} 2009, \pasj, 61, S155

\bibitem[{{Katsuda} {et~al.}(2008){Katsuda}, {Tsunemi}, \&
  {Mori}}]{2008ApJ...678L..35K}
{Katsuda}, S., {Tsunemi}, H., \& {Mori}, K. 2008, \apjl, 678, L35

\bibitem[{{Kothes} {et~al.}(2006){Kothes}, {Fedotov}, {Foster}, \&
  {Uyan{\i}ker}}]{2006A&A...457.1081K}
{Kothes}, R., {Fedotov}, K., {Foster}, T.~J., \& {Uyan{\i}ker}, B. 2006, \aap,
  457, 1081

\bibitem[{{Kothes} {et~al.}(2008){Kothes}, {Landecker}, {Reich}, {Safi-Harb},
  \& {Arzoumanian}}]{2008ApJ...687..516K}
{Kothes}, R., {Landecker}, T.~L., {Reich}, W., {Safi-Harb}, S., \&
  {Arzoumanian}, Z. 2008, \apj, 687, 516

\bibitem[{{Lazendic} {et~al.}(2004){Lazendic}, {Slane}, {Gaensler}, {Reynolds},
  {Plucinsky}, \& {Hughes}}]{2004ApJ...602..271L}
{Lazendic}, J.~S., {Slane}, P.~O., {Gaensler}, B.~M., {et~al.} 2004, \apj, 602,
  271

\bibitem[{{Leahy} \& {Tian}(2006)}]{2006A&A...451..251L}
{Leahy}, D., \& {Tian}, W. 2006, \aap, 451, 251

\bibitem[{{Leahy} \& {Tian}(2007)}]{2007A&A...461.1013L}
{Leahy}, D.~A., \& {Tian}, W.~W. 2007, \aap, 461, 1013

\bibitem[{{Liu} {et~al.}(2017){Liu}, {Bi}, {Lin}, {Wang}, \&
  {Yin}}]{2017PhRvD..96b3006L}
{Liu}, W., {Bi}, X.-J., {Lin}, S.-J., {Wang}, B.-B., \& {Yin}, P.-F. 2017,
  \prd, 96, 023006

\bibitem[{{Liu} {et~al.}(2019){Liu}, {Guo}, \& {Yuan}}]{2019JCAP...10..010L}
{Liu}, W., {Guo}, Y.-Q., \& {Yuan}, Q. 2019, \jcap, 2019, 010

\bibitem[{{Liu} {et~al.}(2020){Liu}, {Lin}, {Hu}, {Guo}, \&
  {Li}}]{2020ApJ...892....6L}
{Liu}, W., {Lin}, S.-j., {Hu}, H.-b., {Guo}, Y.-q., \& {Li}, A.-f. 2020, \apj,
  892, 6

\bibitem[{{Liu} {et~al.}(2018){Liu}, {Yao}, \& {Guo}}]{2018ApJ...869..176L}
{Liu}, W., {Yao}, Y.-h., \& {Guo}, Y.-Q. 2018, \apj, 869, 176

\bibitem[{{Malkov} \& {Moskalenko}(2020)}]{2020arXiv201002826M}
{Malkov}, M.~A., \& {Moskalenko}, I.~V. 2020, arXiv e-prints, arXiv:2010.02826

\bibitem[{{Mavromatakis} {et~al.}(2002){Mavromatakis}, {Boumis},
  {Papamastorakis}, \& {Ventura}}]{2002A&A...388..355M}
{Mavromatakis}, F., {Boumis}, P., {Papamastorakis}, J., \& {Ventura}, J. 2002,
  \aap, 388, 355

\bibitem[{{McComas} {et~al.}(2009){McComas}, {Allegrini}, {Bochsler},
  {Bzowski}, {Christian}, {Crew}, {DeMajistre}, {Fahr}, {Fichtner}, {Frisch},
  {Funsten}, {Fuselier}, {Gloeckler}, {Gruntman}, {Heerikhuisen}, {Izmodenov},
  {Janzen}, {Knappenberger}, {Krimigis}, {Kucharek}, {Lee}, {Livadiotis},
  {Livi}, {MacDowall}, {Mitchell}, {M{\"o}bius}, {Moore}, {Pogorelov},
  {Reisenfeld}, {Roelof}, {Saul}, {Schwadron}, {Valek}, {Vanderspek}, {Wurz},
  \& {Zank}}]{2009Sci...326..959M}
{McComas}, D.~J., {Allegrini}, F., {Bochsler}, P., {et~al.} 2009, Science, 326,
  959

\bibitem[{{Mertsch}(2011)}]{2011JCAP...02..031M}
{Mertsch}, P. 2011, \jcap, 2, 31

\bibitem[{{Miceli} {et~al.}(2008){Miceli}, {Bocchino}, \&
  {Reale}}]{2008ApJ...676.1064M}
{Miceli}, M., {Bocchino}, F., \& {Reale}, F. 2008, \apj, 676, 1064

\bibitem[{{Morlino} {et~al.}(2009){Morlino}, {Amato}, \&
  {Blasi}}]{2009MNRAS.392..240M}
{Morlino}, G., {Amato}, E., \& {Blasi}, P. 2009, \mnras, 392, 240

\bibitem[{Motz {et~al.}(2021)Motz, Link, Adriani, Akaike, \&
  De}]{2021Investigating}
Motz, H., Link, J.~T., Adriani, O., Akaike, Y., \& De, N. G.~A. 2021, in 37th
  International Cosmic Ray Conference

\bibitem[{{Neronov} {et~al.}(2012){Neronov}, {Semikoz}, \&
  {Taylor}}]{2012PhRvL.108e1105N}
{Neronov}, A., {Semikoz}, D.~V., \& {Taylor}, A.~M. 2012, \prl, 108, 051105

\bibitem[{{Panov} {et~al.}(2007){Panov}, {Adams}, {Ahn}, {Batkov},
  {Bashindzhagyan}, {Watts}, {Wefel}, {Wu}, {Ganel}, {Guzik}, {Gunashingha},
  {Zatsepin}, {Isbert}, {Kim}, {Christl}, {Kouznetsov}, {Panasyuk}, {Seo},
  {Sokolskaya}, {Chang}, {Schmidt}, \& {Fazely}}]{2007BRASP..71..494P}
{Panov}, A.~D., {Adams}, J.~H., J., {Ahn}, H.~S., {et~al.} 2007, Bulletin of
  the Russian Academy of Sciences, Physics, 71, 494

\bibitem[{{Panov} {et~al.}(2009){Panov}, {Adams}, {Ahn}, {Bashinzhagyan},
  {Watts}, {Wefel}, {Wu}, {Ganel}, {Guzik}, {Zatsepin}, {Isbert}, {Kim},
  {Christl}, {Kouznetsov}, {Panasyuk}, {Seo}, {Sokolskaya}, {Chang}, {Schmidt},
  \& {Fazely}}]{2009BRASP..73..564P}
{Panov}, A.~D., {Adams}, J.~H., {Ahn}, H.~S., {et~al.} 2009, Bulletin of the
  Russian Academy of Sciences, Physics, 73, 564

\bibitem[{{Perko}(1987)}]{1987A&A...184..119P}
{Perko}, J.~S. 1987, \aap, 184, 119

\bibitem[{{Plucinsky} {et~al.}(1996){Plucinsky}, {Snowden}, {Aschenbach},
  {Egger}, {Edgar}, \& {McCammon}}]{1996ApJ...463..224P}
{Plucinsky}, P.~P., {Snowden}, S.~L., {Aschenbach}, B., {et~al.} 1996, \apj,
  463, 224

\bibitem[{{Qiao} {et~al.}(2019){Qiao}, {Liu}, {Guo}, \&
  {Yuan}}]{2019JCAP...12..007Q}
{Qiao}, B.-Q., {Liu}, W., {Guo}, Y.-Q., \& {Yuan}, Q. 2019, \jcap, 2019, 007

\bibitem[{{Redman} \& {Meaburn}(2005)}]{2005MNRAS.356..969R}
{Redman}, M.~P., \& {Meaburn}, J. 2005, \mnras, 356, 969

\bibitem[{{Reich} {et~al.}(1992){Reich}, {Fuerst}, \&
  {Arnal}}]{1992A&A...256..214R}
{Reich}, W., {Fuerst}, E., \& {Arnal}, E.~M. 1992, \aap, 256, 214

\bibitem[{{Reich} {et~al.}(2014){Reich}, {Sun}, {Reich}, {Gao}, {Xiao}, \&
  {Han}}]{2014A&A...561A..55R}
{Reich}, W., {Sun}, X.~H., {Reich}, P., {et~al.} 2014, \aap, 561, A55

\bibitem[{{Reich} {et~al.}(2003){Reich}, {Zhang}, \&
  {F{\"u}rst}}]{2003A&A...408..961R}
{Reich}, W., {Zhang}, X., \& {F{\"u}rst}, E. 2003, \aap, 408, 961

\bibitem[{{Schwadron} {et~al.}(2014){Schwadron}, {Adams}, {Christian},
  {Desiati}, {Frisch}, {Funsten}, {Jokipii}, {McComas}, {Moebius}, \&
  {Zank}}]{2014Sci...343..988S}
{Schwadron}, N.~A., {Adams}, F.~C., {Christian}, E.~R., {et~al.} 2014, Science,
  343, 988

\bibitem[{{Seo} \& {Ptuskin}(1994)}]{1994ApJ...431..705S}
{Seo}, E.~S., \& {Ptuskin}, V.~S. 1994, \apj, 431, 705

\bibitem[{{Serpico}(2012)}]{2012APh....39....2S}
{Serpico}, P.~D. 2012, Astroparticle Physics, 39, 2

\bibitem[{{Smith} {et~al.}(1994){Smith}, {Cunha}, \&
  {Plez}}]{1994A&A...281L..41S}
{Smith}, V.~V., {Cunha}, K., \& {Plez}, B. 1994, \aap, 281, L41

\bibitem[{{Sun} {et~al.}(2006){Sun}, {Reich}, {Han}, {Reich}, \&
  {Wielebinski}}]{2006A&A...447..937S}
{Sun}, X.~H., {Reich}, W., {Han}, J.~L., {Reich}, P., \& {Wielebinski}, R.
  2006, \aap, 447, 937

\bibitem[{{Sveshnikova} {et~al.}(2013){Sveshnikova}, {Strelnikova}, \&
  {Ptuskin}}]{2013APh....50...33S}
{Sveshnikova}, L.~G., {Strelnikova}, O.~N., \& {Ptuskin}, V.~S. 2013,
  Astroparticle Physics, 50, 33

\bibitem[{Tang {et~al.}(2021)Tang, Xia, Shen, Zu, Feng, Yuan, Fan, \&
  Wu}]{tang2021explanation}
Tang, T.-P., Xia, Z.-Q., Shen, Z.-Q., {et~al.} 2021, Explanation of nearby SNRs
  for primary electron excess and proton spectral bump, arXiv:2109.12496

\bibitem[{{Tang} \& {Piran}(2019)}]{2019MNRAS.484.3491T}
{Tang}, X., \& {Piran}, T. 2019, \mnras, 484, 3491

\bibitem[{{Tian} {et~al.}(2020){Tian}, {Liu}, {Yang}, {Fu}, {Xu}, {Yao}, \&
  {Guo}}]{2020ChPhC..44h5102T}
{Tian}, Z., {Liu}, W., {Yang}, B., {et~al.} 2020, Chinese Physics C, 44, 085102

\bibitem[{{Tomassetti}(2012)}]{2012ApJ...752L..13T}
{Tomassetti}, N. 2012, \apjl, 752, L13

\bibitem[{{Tomassetti}(2015)}]{2015PhRvD..92h1301T}
---. 2015, \prd, 92, 081301

\bibitem[{{Trotta} {et~al.}(2011){Trotta}, {J{\'o}hannesson}, {Moskalenko},
  {Porter}, {Ruiz de Austri}, \& {Strong}}]{2011ApJ...729..106T}
{Trotta}, R., {J{\'o}hannesson}, G., {Moskalenko}, I.~V., {et~al.} 2011, \apj,
  729, 106

\bibitem[{{Vladimirov} {et~al.}(2012){Vladimirov}, {J{\'o}hannesson},
  {Moskalenko}, \& {Porter}}]{2012ApJ...752...68V}
{Vladimirov}, A.~E., {J{\'o}hannesson}, G., {Moskalenko}, I.~V., \& {Porter},
  T.~A. 2012, \apj, 752, 68

\bibitem[{{Xiao} {et~al.}(2009){Xiao}, {Reich}, {F{\"u}rst}, \&
  {Han}}]{2009A&A...503..827X}
{Xiao}, L., {Reich}, W., {F{\"u}rst}, E., \& {Han}, J.~L. 2009, \aap, 503, 827

\bibitem[{{Xu} {et~al.}(2007){Xu}, {Han}, {Sun}, {Reich}, {Xiao}, {Reich}, \&
  {Wielebinski}}]{2007A&A...470..969X}
{Xu}, J.~W., {Han}, J.~L., {Sun}, X.~H., {et~al.} 2007, \aap, 470, 969

\bibitem[{{Yamauchi} {et~al.}(2000){Yamauchi}, {Yokogawa}, {Tomida}, {Koyama},
  \& {Tamura}}]{2000bbxs.conf..567Y}
{Yamauchi}, S., {Yokogawa}, J., {Tomida}, H., {Koyama}, K., \& {Tamura}, K.
  2000, in Broad Band X-ray Spectra of Cosmic Sources, ed. K.~{Makishima},
  L.~{Piro}, \& T.~{Takahashi}, 567

\bibitem[{{Yar-Uyaniker} {et~al.}(2004){Yar-Uyaniker}, {Uyaniker}, \&
  {Kothes}}]{2004ApJ...616..247Y}
{Yar-Uyaniker}, A., {Uyaniker}, B., \& {Kothes}, R. 2004, \apj, 616, 247

\bibitem[{{Yoon} {et~al.}(2017){Yoon}, {Anderson}, {Barrau}, {Conklin},
  {Coutu}, {Derome}, {Han}, {Jeon}, {Kim}, {Kim}, {Lee}, {Lee}, {Lee}, {Lee},
  {Link}, {Menchaca-Rocha}, {Mitchell}, {Mognet}, {Nutter}, {Park},
  {Picot-Clemente}, {Putze}, {Seo}, {Smith}, \& {Wu}}]{2017ApJ...839....5Y}
{Yoon}, Y.~S., {Anderson}, T., {Barrau}, A., {et~al.} 2017, \apj, 839, 5

\bibitem[{{Yuan} \& {Bi}(2013)}]{2013PhLB..727....1Y}
{Yuan}, Q., \& {Bi}, X.-J. 2013, Physics Letters B, 727, 1

\bibitem[{{Yuan} \& {Feng}(2018)}]{2018SCPMA..61j1002Y}
{Yuan}, Q., \& {Feng}, L. 2018, Science China Physics, Mechanics, and
  Astronomy, 61, 101002

\bibitem[{{Yuan} {et~al.}(2020{\natexlab{a}}){Yuan}, {Qiao}, {Guo}, {Fan}, \&
  {Bi}}]{2020arXiv200701768Y}
{Yuan}, Q., {Qiao}, B.-Q., {Guo}, Y.-Q., {Fan}, Y.-Z., \& {Bi}, X.-J.
  2020{\natexlab{a}}, arXiv e-prints, arXiv:2007.01768

\bibitem[{{Yuan} {et~al.}(2020{\natexlab{b}}){Yuan}, {Zhu}, {Bi}, \&
  {Wei}}]{2020JCAP...11..027Y}
{Yuan}, Q., {Zhu}, C.-R., {Bi}, X.-J., \& {Wei}, D.-M. 2020{\natexlab{b}},
  \jcap, 2020, 027

\bibitem[{{Zhang} {et~al.}(2021){Zhang}, {Qiao}, {Liu}, {Cui}, {Yuan}, \&
  {Guo}}]{2021JCAP...05..012Z}
{Zhang}, P.-p., {Qiao}, B.-q., {Liu}, W., {et~al.} 2021, \jcap, 2021, 012

\bibitem[{{Zirnstein} {et~al.}(2016){Zirnstein}, {Heerikhuisen}, {Funsten},
  {Livadiotis}, {McComas}, \& {Pogorelov}}]{2016ApJ...818L..18Z}
{Zirnstein}, E.~J., {Heerikhuisen}, J., {Funsten}, H.~O., {et~al.} 2016, \apjl,
  818, L18

\end{thebibliography}

\end{document}